\documentclass{LMCS}

\usepackage{amsmath,amssymb,latexsym,stmaryrd,mathptm,hyperref}
\usepackage[l>=1em,size=2em]{diagrams}
\usepackage[T1]{fontenc}

\newif\iflong
\longtrue

\makeatletter

\def\@listi{\leftmargin\leftmargini
            \parsep 4\p@ \@plus2\p@ \@minus\p@
            \topsep 4\p@ \@plus2\p@ \@minus\p@
            \itemsep\z@}
\let\@listI\@listi
\@listi

\makeatother

\newtheorem{fact}{Fact}[section]
\newtheorem{lemma}[fact]{Lemma}

\newtheorem{theorem}[fact]{Theorem}
\newtheorem{corollary}[fact]{Corollary}

\newtheorem{proposition}[fact]{Proposition}
\newtheorem{exx}[fact]{Example}
\newtheorem{defii}[fact]{Definition}

\newenvironment{definition}[1][]{%
  \ifx\relax#1\relax
    \begin{defii}\rm
  \else
    \begin{defii}[#1]\rm
  \fi
}{%
  \EndProof
  \global\let\EndProof\EndProofBox
  \end{defii}}

\newenvironment{pf}[1][]{%
  \trivlist\item{\bf Proof\ifx\relax#1\relax:~\else~#1:\fi}%
}{%
  \qed%\EndProof
  \endtrivlist
}

\newcommand{\EndProofBox}{\null\hfill$\Box$}
\global\let\EndProof\EndProofBox
\newcommand{\boxHere}{\global\let\EndProof\empty\EndProofBox}

\newcommand{\hopi}{HO$\mathrel{\pi}$}
\newcommand{\bnf}{\;\;\mid\;\;}
\newcommand{\sep}{\; ; \;}

\newcommand{\nil}{\mathbf{0}}
\newcommand{\Ppar}{\parallel}
\newcommand{\newn}[2]{\nu #1 \mathbin. (#2)}
\newcommand{\newnt}[3]{\nu (#1:#2) \mathbin. (#3)}
\newcommand{\newnop}[2]{\nu #1 \mathbin. #2}
\newcommand{\ifexp}[4]{\ifoneexp{#1 = #2}{#3}{#4}}
\newcommand{\ifoneexp}[3]{\mathsf{if}~#1~\mathsf{then}~#2\ifx\relax#3\relax\else~\mathsf{else}~#3\fi}

\newcommand{\app}[2]{#1 \cdot #2}
\newcommand{\inpsym}{?}
\newcommand{\outsym}{!}
\newcommand{\abst}[2]{(#1)#2}
\newcommand{\conc}[2]{\langle #1 \rangle #2} 
\newcommand{\inp}[3]{#1  \abst{#2}{#3}} 
\newcommand{\out}[3]{#1 \conc{#2}{#3}}
\newcommand{\repl}[1]{* #1}

\newcommand{\names}{\mathcal{N}}
\newcommand{\vars}{\mathcal{V}}
\newcommand{\vx}{x}
\newcommand{\vy}{y}
\newcommand{\vz}{z}
\newcommand{\ca}{a}
\newcommand{\cb}{b}
\newcommand{\Cc}{c}
\newcommand{\iden}{n}
\newcommand{\pp}{P}
\newcommand{\pq}{Q}
\newcommand{\pr}{R}
\newcommand{\typet}{T}
\newcommand{\typeu}{U}

\newcommand{\dd}{d}

\newcommand{\valv}{v}
\newcommand{\valw}{w}

\newcommand{\conf}{C}
\newcommand{\confd}{D}
\newcommand{\confe}{E}
\newcommand{\tnk}{k}
\newcommand{\tnl}{l}

\newcommand{\unit}{\cdot}

\newcommand{\tvarz}{Z}
\newcommand{\tenv}{\Gamma}
\newcommand{\cenv}{\Delta}
\newcommand{\kenv}{\Theta}

\newcommand{\rel}{\mathrel{\mathcal{R}}}
\newcommand{\set}[1]{\left\{ #1 \right\} }
\newcommand{\sem}[1]{[\![ #1 ]\!]}

\newcommand{\openex}[1]{#1^{o}}

\newcommand{\tunit}{\cdot}
\newcommand{\tproc}{\diamond}
\newcommand{\tchan}[1]{\mathsf{ch}[#1]}
\newcommand{\tabs}[1]{#1 \rightarrow \tproc}
\newcommand{\trec}[2]{\mathsf{rec}\,#1. #2}
\newcommand{\tiso}{\sim_{iso}}
\newcommand{\types}{\vdash}
\newcommand{\ttrig}[1]{#1}

\newcommand{\eval}{\mathcal{E}}
\newcommand{\holedot}[1]{\cdot_{#1}}
\newcommand{\hole}{[ \; \cdot \; ]}
\newcommand{\thole}[1]{[ \; \holedot{#1} \; ]}
\newcommand{\ctxt}[1]{C[#1]}

\newcommand{\fn}[1]{\mathsf{fn}(#1)}

\makeatletter
\newenvironment{axioms}
  {\renewcommand{\arraystretch}{1.1}\[\begin{array}{rcll}}
  {\end{array}\]\global\@ignoretrue}

\newenvironment{inferences}
  {\renewcommand{\arraystretch}{2.5}\[\begin{array}{c}}
  {\end{array}\]\global\@ignoretrue}

\makeatother

\newcommand{\infer}[2]{%
  {\renewcommand{\arraystretch}{1}\frac{%
    \begin{array}{c}\displaystyle#1\end{array}%
  }{%
    \begin{array}{c}\displaystyle#2\end{array}%
  }}%
}

\newcommand{\redn}{\lt{}}
\newcommand{\Redn}{\wlt{}}

\newcommand{\trans}[1]{#1\mathrel{}^*}
\newcommand{\hbeta}{\mathsf{h}}
\newcommand{\lthbeta}{\stackrel{\hbeta}{\rightarrow}}
\newcommand{\transhbeta}{\lthbeta\mathrel{}^*}

\newcommand{\sts}{\models}
\newcommand{\convon}[1]{\Downarrow #1}
\newcommand{\ctxteq}{\cong}
\newcommand{\bisim}{\approx}
\newcommand{\parctxteq}{\ctxteq_{p}}
\newcommand{\mgctxtrel}{\bisim_{m}}
\newcommand{\parbisim}{\bisim_{p}}

\newcommand{\call}[1]{\tau_{#1}}
\newcommand{\res}[2]{\langle #1 \Leftarrow #2 \rangle}

\newcommand{\lab}{\alpha}
\newcommand{\taulab}{\tau}
\newcommand{\outlab}[2]{#1 \langle #2 \rangle \outsym}
\newcommand{\inlab}[2]{#1 \langle #2 \rangle \inpsym}
\newcommand{\newoutlab}[2]{\newnop{#2}{\outlab{#1}{\call{#2}}}}
\newcommand{\newinlab}[2]{\newnop{#2}{\inlab{#1}{\call{#2}}}}
\newcommand{\newlab}[2]{\newnop{#1}{#2}}
\newcommand{\dual}[1]{\bar{#1}}

\newarrowtail{<-}\leftarrow\rightarrow\uparrow\downarrow
\newarrow{Reln}{<-}---{->}
\newarrow{Strong}----{->}
\newarrow{Weak}===={=>}
\newarrow{RW}----{>>}

\newcommand{\lt}[1]{\rStrong^{#1}}
\newcommand{\wlt}[1]{\rWeak^{#1}}

\newcommand{\noden}{n}
\newcommand{\nodem}{m}

\newcommand{\rg}[1]{\mathsf{rg}(#1)}
\newcommand{\rgrewrite}{\twoheadrightarrow}
\newcommand{\mg}[1]{\langle \! \langle #1 \rangle \! \rangle}

\newcommand{\ctxtlab}[2]{\mathcal{T}^{#1}_{#2}}

\newcommand{\succbarb}{\delta}
\newcommand{\succsucc}[1]{\out{\succbarb}{#1}{}}
\newcommand{\succout}[1]{(\succsucc{#1} \ic \failout)}
\newcommand{\failbarb}{\delta'}
\newcommand{\failout}{\out{\failbarb}{}{}}
\newcommand{\ic}{\oplus}

\newcommand{\trigger}[1]{Tr_{#1}}
\newcommand{\sanres}[2]{\left\{ #1 := #2 \right\} }

\def\doi{1 (1:4) 2005}
\lmcsheading%
{\doi}
{22}  %% no. of pages
{}
{}
{Sep.~17, 2004} %% only use ``\phantom{0}'' for single digit dates!
{Apr.~21, 2005} %% only use ``\phantom{0}'' for single digit dates!
{}    %% date of revision, if applicable
\begin{document}

\title{Contextual equivalence for higher-order $\pi$-calculus revisited}

\author[A.~Jeffrey]{Alan Jeffrey\rsuper a}
\thanks{{\lsuper a}This material is based upon work supported by the National Science Foundation under Grant No.~0430175}
\address{{\lsuper a}Bell Labs, Lucent Technologies
  \\ and CTI, DePaul University}
\email{ajeffrey@bell-labs.com}

\author[J.~Rathke]{Julian Rathke\rsuper b}
\thanks{{\lsuper b}Research partially funded by the Nuffield Foundation.}
\address{{\lsuper b}School of Informatics \\ University of Sussex}
\email{julianr@sussex.ac.uk}

%\date{Submitted to LMCS~September 2004\\
%  Revised for publication~March 2005}
%\amsclass{}

\begin{abstract}
  The higher-order $\pi$-calculus is an extension of the
  $\pi$-calculus to allow communication of abstractions of processes
  rather than names alone. It has been studied intensively by
  Sangiorgi in his thesis where a characterisation of a
  contextual equivalence for higher-order 
  $\pi$-calculus  is provided using labelled
  transition systems and \emph{normal} bisimulations. 
  Unfortunately the proof technique used there requires a restriction
  of the language to only allow finite types.
   
  We revisit this calculus and offer an alternative presentation of the
  labelled transition system and a novel proof technique which allows us
  to provide a fully abstract characterisation of contextual equivalence 
  using labelled transitions and bisimulations for higher-order
  $\pi$-calculus with recursive types also.
\end{abstract}

\maketitle

%{{{ Introduction
\vskip-\bigskipamount
\section{Introduction}

It is evident that there is growing interest in
the study of mobile code in process languages 
\cite{FGLMD96,CG98:MobileAmbients,RH98:TypedLanguage,VC99:Seal}.
It is also clear that there is some relationship between the 
use of higher-order features and mobility. Indeed, code mobility can
be expressed as communication of process abstractions. 
For this reason then it is important for us to develop a clear
understanding of the use of higher-order features in process
languages.

Work towards this began several years ago with various proposals for
higher-order versions of known calculi \cite{Thom90,GMP89},
including the higher-order $\pi$-calculus or \hopi~\cite{SangiorgiThesis}.
This calculus was studied intensively by Sangiorgi and one of his
achievements was to provide a translation of
the higher-order language which supports code mobility, to a
first-order $\pi$-calculus which supports only name mobility.
This translation is proved to be fully abstract with respect to barbed
congruence, but with the restriction to a language of finite types. 

While the translation is of interest in its own right, it also turned
out to be very useful for providing a powerful
fully abstract characterisation of barbed congruence in terms of
labelled transition systems and \emph{normal} bisimulations. 
Providing direct proof techniques for contextual equivalences in
higher-order process languages is often considered to be hard 
\cite{SangiorgiBook}. 
In this paper, the difficulty arises in establishing soundness of the proof
technique, which is tantamount to establishing some sort of
contextuality property.
It has been seen that the use of a translation of higher- to 
first-order communication can alleviate this problem 
and such translations have been employed to this effect 
\cite{San96a:BisimulationHOPC,JeffreyRathke:tbcmlln}.

However, due to the restriction to finite types for the correctness of
these translations, the soundness of the proof technique is only
guaranteed for finite types. 
Given that recursive types are used extensively in $\pi$-calculus, for
encodings of datatypes and functions, this poses a significant
restriction. 
Sangiorgi has shown that by studying various subcalculi, such as the
asynchronous $\pi$-calculus, he is able to remove the restriction to
finite types \cite{SangiorgiBook}.  To date, there has been no proof
of full abstraction for full \hopi~in the presence of recursive types.

In this paper we present
an alternative description of labelled transition systems and normal
bisimulations for \hopi, which is informed by Sangiorgi's
translation of higher-order to first-order communication.
Our alternative presentation allows a 
\emph{direct} proof of soundness for contextual equivalence which
makes no use of the translation to first-order $\pi$-calculus and,
more importantly, makes no restriction on types.

The innovation here lies in the introduction of operators 
$\call{\tnk}$ and $\res{\tnk}{\valv}$
which simulate the triggers $\trigger{\tnk}$ and meta-notation 
$\sanres{\tnk}{\valv}$ of Sangiorgi \cite{San96a:BisimulationHOPC} where $\tnk$
is a unique identifier for the trigger and $\valv$ is a process abstraction.
The crucial difference is that where Sangiorgi gives definitions as 
\hopi~terms for these devices:
\[
  \trigger{\tnk} =  \abst{\vx}{\out{\tnk}{\vx}{}} \qquad \mbox{and}
  \qquad \sanres{\tnk}{\valv} = \repl{ \inp{\tnk}{\vx}{\app{\valv}{\vx}}}
\]
where $\out{\tnk}{\vx}{}$ represents an output on name $\tnk$ and 
$\repl{\inp{\tnk}{\vx}{\pp}}$ represents a replicated input on name
  $\tnk$,
we leave the operators uninterpreted. There are no interactions between the
operators $\call{\tnk}$ and $\res{\tnk}{\valv}$. Rather, we just mimic
the behaviour of triggers in the labelled transition systems. The benefit of
doing this is that it allows us to obtain a direct soundness proof that
(normal) bisimilarity implies contextual equivalence without recourse
to any translation in its correctness proof.

A challenge of approaching the problem in this way is that
it is not immediately clear that bisimilarity will be complete for
contextual equivalence in \hopi. That is to say, it is not obvious 
whether each transition has a genuine \hopi~context which validates
it. At this point however we can interpret the operators $\call{\tnk}$
and $\res{\tnk}{\valv}$ as \hopi~terms exactly as Sangiorgi does. 
It is then a simple matter to demonstrate completeness following
familiar techniques 
\cite{FGLMD96,JeffreyRathke:tbcmlln,HR:tbepps02}.
The real payoff is that not only do we obtain a direct soundness proof
but the postponement of interpreting the triggers allows us to finesse any
restrictions to finite types. 

The remainder of the paper is organised as follows: in Section~2 we
recall the syntax and semantics of \hopi~along with the definition of
contextual equivalence which we will be using. This is followed in
Section~3 by a presentation of the novel labelled transition system
using the operators $\call{\tnk}$ and $\res{\tnk}{\valv}$. We prove
that bisimilarity over this labelled transition system is sound for
contextual equivalence in Section~4 and conversely, that it is
complete for contextual equivalence in Section~5. We conclude in
Section~6 with some closing remarks.

\iflong\else
  In this extended abstract, we elide some relatively routine proofs.
  Since much of the novelty of this paper is in our technique
  for establishing soundness, we provide all of the proofs in Section~4.
\fi

%}}}
%{{{ language and equivalence

\section{Higher-order $\pi$ calculus}

Except for
small changes in notation the language is as can be found in 
\cite{SangiorgiBook} with three main differences: 
\begin{enumerate}

\item We assume two distinct countably infinite sets of
  identifiers, $\vars$ and $\names$, 
  for variables and channel names respectively. 
  In general we will use $\vx, \vy, \vz$ to range over variables 
  and $\ca, \cb, \Cc$ to range over channel
  names. This variable/name distinction makes the algebraic
  properties of the language a little cleaner and we are confident that
  the techniques proposed here would also be applicable if we identified
  these sets. 

\item Since we have adopted a variable/name distinction,
  we have used Honda and Yoshida's definition of
  observational equivalence~\cite{HondaYoshida:Redbps}
  in Section~\ref{Sec:contextual-equivalence} rather than Sangiorgi's.
  See~\cite{FournetGonthier:hieeac} for a discussion of this issue.

\item We allow communication of channel names as well
  as process abstractions so that there is a core $\pi$-calculus as a
  direct subcalculus of \hopi.

\end{enumerate}

\subsection{Syntax}

We present the syntax of \hopi~in Figure~\ref{fig:syntax}. 
%
%{{{ grammar
%
%
\begin{figure}
\begin{axioms}
\typet, \; \typeu & ::= &         & \textbf{Value Types}\\
      &  & \tunit                 & \text{Unit type} \\
      &  & \tchan{\typet}         & \text{Channel type} \\
      &  & \tabs{\typet}          & \text{Abstraction type} \\
      &  & \tvarz                 & \text{Type variable} \\
      &  & \trec{\tvarz}{\typet}  & \text{Recursive type} \\
\\
\pp,\;\pq &::=&                   &\textbf{Terms}\\
      &  & \app{\valv}{\valw}     &\text{Application} \\
      &  & \inp{\valv}{\vx:\typet}{\pp}     &\text{Input}\\
      &  & \out{\valv}{\valw}{\pp}         &\text{Output}\\
      &  & \ifexp{\valv}{\valw}{\pp}{\pq} &\text{Matching}\\
      &  & \newnt{\ca}{\typet}{\pp}     &\text{Name creation}\\
      &  & \pp \Ppar \pq          &\text{Concurrency} \\
      &  & \repl{\pp}               &\text{Repetition} \\
      &  & \nil              &\text{Termination}\\
      \\  
\valv,\;\valw &::=&                     &\textbf{Values}\\
      &   & \unit                 &\text{Unit value} \\
      &   & \ca                   &\text{Channel name}\\
      &   & \vx                   &\text{Variable}\\
      &   & \abst{\vx : \typet}{\pp}       &\text{Abstractions}
\end{axioms}

\caption{The Syntax\label{fig:syntax}}
\end{figure} 
%
%}}}
%
The grammar of types for values includes:
\begin{itemize}

\item $(\tunit)$: a singleton type just containing the value
  $(\unit)$.

\item $\tchan{\typet}$: the type of channels which can
  be used for communicating data of type $\typet$.
  Note that in this paper we are not considering input-only
  or output-only channels.

\item $\tabs{\typet}$: the type of an abstraction $\abst{\vx:\typet}{\pp}$.
  Such an abstraction can be applied to a value $\valv$ of type $\typet$
  to return a well-typed process $\pp[\valv/\vx]$.

\item $\tvarz$ and $\trec{\tvarz}{\typet}$:
  these allow recursive types, such as the type
  for monomorphic $\pi$-calculus channels
  $\trec{\tvarz}{\tchan{\tvarz}}$.  We require
  $\tvarz$ to be \emph{guarded}:
  any free occurrence of $\tvarz$
  lies within a subexpression of $\typet$ of the form $\tchan{\typeu}$
  or $\tabs{\typeu}$. 

\end{itemize}
The grammar of process terms includes:
\begin{itemize}

\item $\app{\valv}{\valw}$: the application of abstraction
  $\valv$ to argument $\valw$.  During execution,
  $\valv$ will be instantiated by an abstraction of the
  form $\abst{\vx:\typet}{\pp}$, and $\beta$-reduction will
  give the process $\pp[\valw/\vx]$.

\item $ \inp{\valv}{\vx:\typet}{\pp}$ and $\out{\valv}{\valw}{\pp}$,
  which are the standard synchronous input and output of the $\pi$-calculus,
  except that since abstractions are first-class values,
  we can communicate higher-order data as well as first-order data.

\item $\ifexp{\valv}{\valw}{\pp}{\pq}$:
  an equality test on values, where the type system will ensure
  that $\valv$ and $\valw$ are channels, and so we will never
  compare abstractions for syntactic identity.

\item $\newnt{\ca}{\typet}{\pp}$, $\pp \Ppar \pq$, $\repl{\pp}$ and $\nil$:
  the standard $\pi$-calculus processes for channel generation,
  concurrency, replication and termination.

\end{itemize}
The grammar of values includes:
\begin{itemize}

\item $(\unit)$: the only value of type $(\tunit)$.

\item $\ca$ and $\vx$: channel names and variables respectively.

\item $\abst{\vx:\typet}{\pp}$: an abstraction, which can be
  applied to a value $\valv$ to return a process $\pp[\valv/\vx]$.
  Since abstractions are considered first-class values, they
  can be communicated on channels, or passed as arguments to
  other abstractions.  This feature gives \hopi~its higher-order
  power.

\end{itemize}

\subsection{Reduction semantics}

The reduction semantics for the language is defined in a standard
manner: we first introduce the evaluation contexts
$$
\eval ::= \hole \bnf \eval \Ppar \pp \bnf \newnop{\ca}{\eval}$$ 
Structural equivalence, $\equiv$ is defined to be the least congruence with
respect to $\eval$ contexts such that it makes $(\Ppar, \nil)$ into a
commutative monoid and moreover satisfies
\begin{axioms}
\newn{\ca}{\pp \Ppar \pq} & \equiv & \newnop{\ca}{\pp} \Ppar \pq
\qquad &
\mbox{if } \ca \not\in \fn{\pp}
\\
\repl{\pp} & \equiv & \repl{\pp} \Ppar \pp
\end{axioms}
We will now consider processes up to structural equivalence throughout
the remainder. We define the reduction relation $\redn$ as the least
precongruence with respect to $\eval$ contexts such that the following
axioms hold
$$
\begin{array}{llclr}
(\mbox{comm}) & \out{\ca}{\valv}{\pp} \Ppar \inp{\ca}{\vx}{\pq} & \redn & \pp \Ppar
\app{\abst{\vx}{\pq}}{\valv} & \\
(\beta-\mbox{redn}) & \app{\abst{\vx}{\pp}}{\valv} & \redn & \pp[\valv / \vx] & \\
(\mbox{cond---tt}) \qquad & \ifexp{\ca}{\ca}{\pp}{\pq} & \redn & \pp & \\
(\mbox{cond---ff}) & \ifexp{\ca}{\cb}{\pp}{\pq} & \redn & \pq \quad & (\ca \neq \cb)
\end{array} 
$$
In a standard notation we write $\Redn$ to denote the reflexive,
transitive closure of $\redn$.
% and we sometimes write $\lt{\beta}$
%to signify that axiom $\beta$-redn was used to derive a reduction and  
%$\transbeta$ for the reflexive transitive closure of $\lt{\beta}$.

\subsection{Type system}

We introduce a simple type system for the language which comprises
types for channels and abstractions, together with recursive types.
%
%We also allow recursive types of
%the form $\trec{\tvarz}{\typet}$ where $\trec{}{}$ forms a binder and
%$\tvarz$ is drawn from a countably infinite supply of type variables.
%We must insist that for any $\trec{\tvarz}{\typet}$ that 
%$\tvarz$ does not appear unguarded in $\typet$, that is to say that
%any free occurrence of $\tvarz$
%lies within a subexpression of $\typet$ of the form $\tchan{\typeu}$
%or $\tabs{\typeu}$. 
%
To allow us to infer recursive types for terms we make use of type
isomorphism. We define this by letting $\tiso$ be the least
congruence on types which includes 
$$
\trec{\tvarz}{\typet} \tiso \typet[ \trec{\tvarz}{\typet} / \tvarz ]
$$
A type environment $\tenv$ is a finite set of mappings from
identifiers (channel names or variables) to types
with the restriction that channel names $\ca$ must be mapped to
channel types of the form $\tchan{\typet}$.
We write $\tenv,
\iden : \typet$ to represent the environment made up of the disjoint
union of $\tenv$ and the mapping $\iden$ to $\typet$. We will call an
environment \emph{closed} if it contains mappings of channel names
only and will write $\cenv$ to indicate this. Type inference rules for
the calculus are given in Figure~\ref{fig:type-rules}. We will call a
well-typed process, $\pp$, closed if it can be typed as 
$\Delta \types \pp$ for some closed $\Delta$.
It is easily shown that subject reduction holds for closed terms 
for the reduction relation and type inference system given. 

%{{{ type rules

\begin{figure}
\begin{inferences}
  \infer{}
      {\tenv \types \unit : \tunit}
     \qquad
  \infer{\tenv (\valv) = \typet}
      {\tenv \types \valv : \typet}
     \qquad
 \infer{\tenv, \vx : \typet \types \pp}
    {\tenv \types \abst{\vx : \typet}{\pp} : \tabs{\typet} }
    \qquad 
%    \infer{\tenv \types \valv : \typet[\trec{\tvarz}{\typet} /
%      \tvarz]}
%    {\tenv \types \valv : \trec{\tvarz}{\typet}}   
    \infer{\tenv \types \valv : \typet \quad \typet \tiso \typeu}
    {\tenv \types \valv : \typeu}
    \\  \\
    \infer{ 
      \tenv \types \valv : \tchan{\typet}, \valw : \tchan{\typet}
      \\
      \tenv \types \pp
      \qquad
      \tenv \types \pq
      }
    {\tenv \types  \ifexp{\valv}{\valw}{\pp}{\pq}}
    \qquad   
    \infer{\tenv, \ca :\typet \types  \pp}
    {\tenv \types \newnt{\ca}{\typet}{\pp}}
    \qquad
    \infer{\tenv\types \pp,\; \pq}
    {\tenv \types \pp \Ppar \pq,\; \repl{\pp},\; \nil} 
    \\[\bigskipamount]
    \infer{
      \tenv \types \valv : \tabs{\typet} \quad \tenv \types \valw :
      \typet}
    {\tenv \types \app{\valv}{\valw}}
    \quad 
 \infer    { \tenv, \vx:\typet \types \pp
      \quad
      \tenv \types \valv : \tchan{\typet}}
{\tenv \types \inp{\valv}{\vx:\typet}{\pp}}
    \quad
    \infer    {  \tenv \types \pp
      \quad
      \tenv \types  \valw: \typet
      \quad
      \tenv \types \valv : \tchan{\typet}
    }
{\tenv \types \out{\valv}{\valw}{\pp}}   
\end{inferences}

\caption{The Typing Rules\label{fig:type-rules}}
\end{figure}

%}}}

\subsection{Contextual equivalence}
\label{Sec:contextual-equivalence}

We will now define an appropriate notion of behavioural equivalence
based on contexts and barbs.

Contexts are defined by extending the syntax of processes by allowing
typed holes $\thole{\tenv}$ in terms. The type inference system is
extended to contexts by using the rule
\begin{inferences}
\infer{}{\tenv, \tenv' \types \thole{\tenv}}
\end{inferences}
We write $\ctxt{}$ to denote contexts with at most one hole and
$\ctxt{\pp}$ for the term which results from substituting $\pp$ into
the hole. 

For any given channel name $\ca$ such that $\cenv \types a :
\tchan{\tunit}$ we write $\cenv \sts \pp \convon{\ca}$ if there exists
some $\pp',\pp''$ such that  $\pp \Redn \newn{\cenv'}{
  \out{\ca}{\unit}{\pp''} \Ppar \pp'}$ with $\ca\not\in\cenv'$. 

We use type-indexed families of relations $\set{\rel_\cenv}$ between
closed process terms to describe equivalence. We will write $\rel$ to
refer to the whole family of relations and  
$$
\cenv \sts \pp \rel \pq
$$
to indicate that $\pp$ and $\pq$ are well-typed with respect to
$\cenv$ and related by $\rel_\cenv$.
For general process terms we define the \emph{open extension}
$\openex{\rel}$ of a
typed relation $\rel$ as 
$$
\cenv, \vx_1 : \typet_1, \ldots , \vx_n : \typet_n \sts \pp \openex{\rel} \pq
$$
holds if for every $\cenv'$ disjoint from $\cenv$ and every
$\valv_i$ such that $\cenv, \cenv' \types \valv_i : \typet_i$ (for $1
\leq i \leq n$)
we have
\[
  \cenv, \cenv' \sts \pp[ \valv_1, \ldots , \valv_n / \vx_1,
    \ldots , \vx_n] \rel \pq[ \valv_1, \ldots , \valv_n / \vx_1, \ldots ,\vx_n]
\]
Note that, in general, for closed terms $\cenv \sts \pp \rel \pq$ is not
equivalent to $\cenv \sts \pp \openex{\rel} \pq$ as
$\openex{\rel}$ enjoys the weakening property that
$\cenv, \cenv' \sts \pp \openex{\rel} \pq$ whenever $\cenv \sts \pp
\openex{\rel} \pq$, even when $\rel$ does not.
However, the contextual equivalence which we study in
this paper is defined as an open extension and therefore will
satisfy this weakening.

There are a number of properties of type-indexed relations that we must
define:

\begin{description}

\item[Symmetry:]
A type-indexed relation $\rel$ is symmetric whenever $\cenv \sts \pp \rel
\pq$ implies $\cenv \sts \pq \rel \pp$.

\item[Reduction closure:]
A type-indexed relation $\rel$ is reduction-closed whenever $\cenv \sts \pp
\rel \pq$ and $\pp \redn \pp'$ implies there exists some $\pq'$ such
that $\pq \Redn \pq'$ and $\cenv \sts \pp' \rel \pq'$.

\item[Contextuality:]
A type-indexed relation $\rel$ is contextual whenever
$\tenv' \sts \pp \openex{\rel} \pq$ and 
$\tenv \types \ctxt{\holedot{{\tenv'}}}$ implies $\tenv \sts \ctxt{\pp}
\openex{\rel} \ctxt {\pq}$.

\item[Barb preservation:]
A type-indexed relation $\rel$ is barb-preserving
if $\cenv \sts \pp \rel \pq$ and $\cenv \sts \pp \convon{\ca}$ implies
$\cenv \sts \pq \convon{\ca}$.

\end{description}

\begin{definition}[Contextual equivalence]
Let $\ctxteq$ be the open extension of the largest type-indexed
relation which is symmetric, reduction-closed, contextual and barb-preserving.
\end{definition}
For technical convenience it will be useful to work with a lighter
definition of contextuality. We say that a relation $\rel$ is
$\Ppar$-contextual if it is preserved by all contexts of the form
$\thole{\tenv} \Ppar \pr$ and we let $\parctxteq$ denote the open
extension of the largest
typed relation over processes which is symmetric, $\Ppar$-contextual,
reduction-closed and barb-preserving.
The following lemma demonstrates that this lighter definition 
is sufficient.
\begin{lemma}[Context lemma] \label{lemma:context-lemma}
$\tenv \sts \pp \ctxteq \pq \quad \mbox{if and only if} \quad 
\tenv \sts \pp \parctxteq \pq$
\end{lemma}

\iflong 
\proof%\begin{pf}
  In Appendix~\ref{app:context-lemma}.
\qed%\end{pf}

\fi

%}}}
%{{{ labelled transitions

\section{Full abstraction}

In this section, we will present a bisimulation equivalence for
\hopi, and show that this equivalence is fully abstract for
contextual equivalence.  

\subsection{Labelled transitions}

We will use a labelled transition system to characterize $\ctxteq$
over higher-order $\pi$-calculus terms. The style of the labelled
transition system differs a little from previous transition systems
offered for \hopi. Most notably, the nodes of the transition
system are described using an augmented syntax rather than process
terms alone. Specifically, for each $\tnk$ drawn from a countable 
set of names disjoint from $\names$ and $\vars$, we introduce two new operators:
$$\call{\tnk} \qquad  \mbox{and} \qquad \res{\tnk}{\valv}$$
with the intuitive reading that $\call{\tnk}$ is an indirect reference to an
abstraction and $\res{\tnk}{\valv}$ stores the abstraction to which $\tnk$
refers so that access to $\valv$ is provided through interaction with $\tnk$.
The augmented syntax for nodes is given the grammar of configurations
$\conf$ obtained by extending
Figure~\ref{fig:syntax} with:
\begin{axioms}
\valv & ::= & \ldots \mbox{(as Figure~\ref{fig:syntax})} \ldots \bnf \call{\tnk} & \\
\conf & ::= & \pp \bnf \res{\tnk}{\valv} \bnf \newn{\ca : \typet}{\conf} \bnf
\conf \Ppar \conf  
\end{axioms}
%where $\tnk$ ranges over a countable set of reference names disjoint from $\names$
%and $\vars$. 
We impose a syntactic restriction on the augmented syntax so that in
any configuration $\conf$ for any given $\tnk$ then
%either 
%\begin{itemize}
%\item $\tnk$ does not appear in $\conf$ or
%\item $\call{\tnk}$ appears in $\conf$ (possibly many times) or
%\item 
$\res{\tnk}{\valv}$ appears at most once in $\conf$.
%\end{itemize}
Structural equivalence and reduction lift to $\conf$ in
the obvious manner ---
note that there are no reduction rules given for $\call{\tnk}$ and
$\res{\tnk}{\valv}$ though.
We augment the type rules by considering judgements of the form 
$$
\tenv \sep \kenv \types \valv : \typet \qquad \mbox{and} \qquad \tenv \sep \kenv
\types \conf
$$ 
where $\kenv$ represents 
%an \emph{ordered} 
a set of mappings from
reference names to types 
%of the form 
$\ttrig{\typet}$.
The rules in Figure~\ref{fig:type-rules} are easily decorated with the
extra $\kenv$ environment. The further rules required are given by
\begin{inferences}
%\infer{\kenv (\tnk) = \tchan{\typet}}{\tenv \sep \kenv \types \tnk :
%  \tchan{\typet}}
%\qquad
\infer{\kenv (\tnk) = \ttrig{\typet}}{\tenv \sep \kenv
  \types \call{\tnk} : \tabs{\typet}}
\qquad
\infer{\kenv (\tnk) = \ttrig{\typet} \quad \tenv \sep
  \kenv \types \valv : \tabs{\typet}}{\tenv \sep \kenv \types \res{\tnk}{\valv}}
\end{inferences}
Nodes of our labelled transition system then are well-typed closed
terms of the augmented language of the form
$$
(\cenv \sep \kenv \types \conf)
$$
The transitions are of the form
$(\cenv\sep\kenv\types\conf)\lt{\lab}(\cenv\sep\kenv\types\conf)$
or
$(\cenv\sep\kenv\types\conf)\lt{\taulab}(\cenv\sep\kenv\types\conf)$
where visible labels $\lab$ are given by the grammar:
\begin{eqnarray*}
  \lab & ::= & \newlab{\ca}{\lab}
    \bnf \newoutlab{\dd}{\tnk}
    \bnf \newinlab{\dd}{\tnk}
    \bnf \inlab{\dd}{\valv}
    \bnf \outlab{\dd}{\valv}
\end{eqnarray*}
where write $\dd$ to mean either a channel name $\ca$ or an indirect
reference name $\tnk$. 
The transitions are presented in 
Figures~\ref{fig:lts1},\ref{fig:lts2},\ref{fig:lts3}.
The intuition for these transitions is (eliding types for readability):
\begin{itemize}

\item $P \lt{\inlab{\ca}{\valv}} P'$:
  indicates that $P$ is prepared to input a value $\valv$ on channel
  $\ca$ and then perform as $P'$.  The type system enforces
  that $\valv$ is a first-order value, and not an abstraction.
  Moreover, in this case both $\ca$ and $\valv$ are pre-existing
  values, and were not generated fresh for this transition.

\item $P \lt{\inlab{\tnk}{\valv}} P'$: 
  indicates that $P$ has provided a named abstraction reference
  $\tnk$ to the environment, and that the environment is calling
  the abstraction with pre-existing argument $\valv$.

\item $P \lt{\newlab{\cb}{\inlab{\ca}{\cb}}} P'$:
  indicates that $P$ is prepared to input a fresh channel
  $\cb$ on channel $\ca$ and then perform as $P'$.
  This is the same as $P \lt{\inlab{\ca}{\cb}} P'$,
  except that $\cb$ is now a fresh channel generated by
  the environment, and has not been seen before by the process.

\item $P \lt{\newlab{\cb}\inlab{\tnk}{\cb}} P'$: 
  indicates that $P$ has provided a named abstraction reference
  $\tnk$ to the environment, and that the environment is calling
  the abstraction with fresh argument $\cb$.

\item $P \lt{\newinlab{\ca}{\tnl}} P'$:
  indicates that $P$ is prepared to input an abstraction
  $\tnl$ on channel $\ca$ and then perform as $P'$.
  In this case, we do not record the abstraction itself
  in the label, but instead we just generate a fresh
  reference $\tnl$ to the abstraction.

\item $P \lt{\newinlab{\tnk}{\tnl}} P'$:
  indicates that $P$ has provided a named abstraction reference
  $\tnk$ to the environment, and that the environment is calling
  that abstraction with argument $\tnl$.  In this case,
  $\tnk$ must be a higher-order abstraction, so is expecting
  an abstraction as an argument.  Rather than recording
  the abstraction itself in the label, we instead generate
  a fresh reference $\tnl$ to the abstraction.

\item Each of the above input transitions has a dual output transition,
  where the role of the process and environment are exchanged.

\end{itemize}
%We write $\wlt{\lab}$ to mean the composite relation $\wktran \lt{\lab}
%\wktran$ which describes the \emph{weak} transitions.
We write $\dual{\lab}$ to denote the complement of an action $\lab$,
which is defined to be the action $\lab$ with the input/output
annotation inversed. We will often write $\Redn$ to mean the reflexive
transitive closure of $\lt{\taulab}$ and $\wlt{\lab}$ to mean 
$\Redn \lt{\lab} \Redn$. 
%$$
%\begin{array}{lcl}
%\dual{\taulab} & = & \taulab \\
%\dual{\gamma \inpsym} & = & \gamma \outsym \\
%\dual{\gamma \outsym} & = & \gamma \inpsym
%\end{array}
%$$
The following proposition states that the labelled transition system
is well-defined in the sense that the transition relation only relates 
well-typed terms.  
\begin{proposition}
  If $\cenv \sep \kenv \types \conf$ and $(\cenv \sep \kenv \types
  \conf) \lt{\lab} (\cenv, \cenv' \sep \kenv, \kenv' \types \conf')$
then $\cenv, \cenv' \sep \kenv, \kenv' \types \conf'$ is a valid
typing judgement.
\end{proposition}
\proof%\begin{pf}
  Straightforward induction.
\qed%\end{pf}

\subsection{Bisimilarity}

We use a standard definition of (weak) bisimilarity to provide our
characterisation of $\mathrel{\ctxteq}$ for \hopi:
\begin{definition}
We call a symmetric relation,
$\rel$,  between nodes of the labelled transition system a
\emph{bisimulation} if whenever $(\noden, \nodem) \in \rel$ we have
\begin{itemize}
 \item $\noden \lt{\taulab} \noden'$  implies there exists some $\nodem'$ such
 that $\nodem \Redn \nodem'$ and $(\noden', \nodem') \in \rel$ 
 \item $\noden \lt{\lab} \noden'$ implies there exists some $\nodem'$ such
 that $\nodem \wlt{\lab} \nodem'$ and $(\noden', \nodem') \in \rel$ 
\end{itemize}
Let bisimulation equivalence, or bisimilarity, $\bisim$ be the largest
bisimulation relation. 
\end{definition}
We will write 
$$ \cenv \sep \kenv \sts \conf \bisim \confd $$
to mean that $\cenv \sep \kenv \types \conf$ and $\cenv \sep \kenv
\types \confd$ are valid typing judgements and moreover, they are
related by $\bisim$ as nodes of the lts. 
In order to provide a bisimulation characterisation of
$\mathrel{\ctxteq}$ over \hopi~we will consider a subrelation of 
$\bisim$ by restricting our attention to nodes of the form
$$
(\cenv \sep \types \pp)
$$
whose terms are clearly definable in \hopi. 
We will simply write (when $\kenv$ is empty)
$$
\cenv \sts \pp \bisim \pq
$$
to indicate bisimilarity between such terms of \hopi~considered as
nodes of the labelled transition system.
%We use the open extension $\mathrel{\openex{\bisim}}$ as defined above
%to relate open \hopi~terms:
%%This relation is only defined for closed terms so far and we lift it
%%to open terms by 
%let
%$$
%\cenv , \vx_1: \typet_1 , \ldots , \vx_n : \typet_n \sts \pp \openex{\bisim} \pq
%$$
%hold if and only if 
%for every $\cenv'$ disjoint from $\cenv$ and every
%$\valv_i$ of \hopi~such that $\cenv, \cenv' \types \valv_i : \typet_i$ 
%(for $1 \leq i \leq n$)
%we have $\cenv, \cenv'  \sts \pp[ \valv_1, \ldots , \valv_n / \vx_1,
%\ldots , \vx_n] \rel \pq[ \valv_1, \ldots , \valv_n / \vx_1, \ldots
%,\vx_n]$.
%Note though that these values $v_i$ may contain indirect reference
%values $\call{\tnk}$.
%It is also useful to define equivalence between two values. For
%non-abstraction values this is simply the identity relation but for
%abstractions 
%we have 
%$$
%\tenv \sep \kenv \sts \abst{\vx : \typet}{\pp} \openex{\bisim} \abst{\vx:\typet}{\pq}
%$$
%if and only if
%$\tenv , \vx : \typet \sep \kenv \sts \pp \openex{\bisim} \pq$.

%It is clear that $\openex{\bisim}$ forms a symmetric, reduction
%closed and barb-preserving relation. 
%By Lemma~\ref{lemma:context-lemma}, we only need to show that 
%it is preserved by $\Ppar$-contexts over \hopi.
%We consider bisimilarity on {\hopi} as an instance of the above relation
%between well-typed terms of the augmented syntax
%by defining, for well-typed terms $\pp, \pq$ of \hopi
%$$
%\cenv \sts \pp \bisim \pq \qquad \mbox{if and only if} \qquad \cenv
%\sep \sts \pp \bisim \pq
%$$

%{{{ lts

\begin{figure}
  \begin{inferences}
    \infer{\conf \redn \conf'}
    {(\cenv \sep \kenv \types \conf) \lt{\taulab} (\cenv \sep \kenv \types
      \conf')}
\qquad
    \infer{(\cenv \sep \kenv \types \conf) \lt{\lab} (\cenv' \sep
      \kenv' \types \conf')}
    {(\cenv \sep \kenv \types \conf \Ppar \confd) \lt{\lab} (\cenv'
      \sep \kenv' \types \conf' \Ppar \confd)}
\\[\bigskipamount]
    \infer{(\cenv, \ca : \typet \sep \kenv \types \conf) \lt{\lab}
           (\cenv, \ca : \typet, \cenv' \sep \kenv, \kenv' \types
           \conf') \quad (\ca \not\in \fn{\lab})}
         {(\cenv \sep \kenv \types \newnop{\ca : \typet}{\conf})
           \lt{\lab}
           (\cenv, \cenv' \sep \kenv, \kenv' \types \newnop{\ca:
             \typet}{\conf'})}
\\[\bigskipamount]
    \infer{(\cenv, \cb : \typet \sep \kenv \types \conf) \lt{\outlab{\dd}{\cb}}
      (\cenv, \cb : \typet \sep \kenv \types \conf') \quad (\dd \neq \cb)}
    {(\cenv \sep \kenv \types \newnop{\cb : \typet}{\conf})
      \lt{\newlab{\cb}{\outlab{\dd}{\cb}}}
     (\cenv, \cb : \typet \sep \kenv \types \conf')}
\\[\bigskipamount]
\infer{(\cenv, \cb: \typet \sep \kenv \types \conf)
  \lt{\inlab{\dd}{\cb}} (\cenv, \cb : \typet \sep \kenv \types
  \conf') \quad (\dd \neq \cb)}
{(\cenv \sep \kenv \types \conf) \lt{\newlab{\cb}{\inlab{\dd}{\cb}}} (\cenv, \cb
  : \typet \sep \kenv \types \conf')}
\end{inferences}
\caption{Structural labelled transition rules}\label{fig:lts1}
\end{figure}

\begin{figure}
\begin{inferences}
\infer{\typet \tiso \tabs{\typeu}}{(\cenv \sep \kenv \types \inp{\ca}{\vx : \typet}{\pp})
  \lt{\newinlab{\ca}{\tnk}}
       (\cenv \sep \kenv, \tnk:\ttrig{\typeu} \types
       \app{\abst{\vx:\typet}{\pp}}{\call{\tnk}})}
\\[\bigskipamount]
\infer{\kenv(\tnk) \tiso \tabs{\typet}}{
(\cenv \sep \kenv \types \res{\tnk}{\valv}) \lt{\newinlab{\tnk}{\tnl}}
(\cenv \sep \kenv, \tnl : \ttrig{\typet} \types
\app{\valv}{\call{\tnl}} \Ppar \res{\tnk}{\valv})}
\\[\bigskipamount]
\infer{\cenv \sep \kenv \types \valv : \tabs{\typet}}
{(\cenv \sep \kenv \types \out{\ca}{\valv}{\pp})
  \lt{\newoutlab{\ca}{\tnk}}
(\cenv \sep \kenv, \tnk: \ttrig{\typet} \types \res{\tnk}{\valv} \Ppar
\pp)}
\\[\bigskipamount]
\infer{\kenv(\tnk) \tiso \tabs{\typet}}
{(\cenv \sep \kenv \types \app{\call{\tnk}}{\valv})
  \lt{\newoutlab{\tnk}{\tnl}}
(\cenv \sep \kenv, \tnl: \ttrig{\typet} \types \res{\tnl}{\valv})}
  \end{inferences}
    \caption{Basic higher-order labelled transition rules}
    \label{fig:lts2}
\end{figure}

\begin{figure}
  \begin{inferences}
    \infer{\cenv \types \valv : \typet \mbox{ a base type}}
    {(\cenv \sep \kenv \types \inp{\ca}{\vx : \typet}{P})
  \lt{\inlab{\ca}{\valv}}
       (\cenv \sep \kenv \types
       \app{\abst{\vx:\typet}{\pp}}{\valv})}
\\[\bigskipamount]
\infer{\kenv(\tnk) = \ttrig{\typet} \quad \cenv \types \valw : \typet
  \mbox{ a base type}}{
(\cenv \sep \kenv \types \res{\tnk}{\valv}) \lt{\inlab{\tnk}{\valw}}
(\cenv \sep \kenv \types
\app{\valv}{\valw} \Ppar \res{\tnk}{\valv})}
\\[\bigskipamount]
\infer{\cenv \types \valv : \typet \mbox{ a base type}}
{(\cenv \sep \kenv \types \out{\ca}{\valv}{\pp})
  \lt{\outlab{\ca}{\valv}}
(\cenv \sep \kenv \types \pp)}
\\[\bigskipamount]
\infer{\kenv(\tnk) = \ttrig{\typet} \quad \typet \mbox{ a base type}}
{(\cenv \sep \kenv \types \app{\call{\tnk}}{\valv})
  \lt{\outlab{\tnk}{\valv}}
(\cenv \sep \kenv \types \nil)}
  \end{inferences}
  \caption{Basic first-order labelled transition rules}
  \label{fig:lts3}
\end{figure}

%}}}

%}}}
%{{{ Congruence

\subsection{Soundness of bisimilarity for contextual equivalence}

We need to demonstrate that bisimilarity 
implies contextual equivalence for all \hopi~processes. 
In particular, because of Lemma~\ref{lemma:context-lemma}, we need
only show that
bisimilarity is contained in some symmetric, reduction-closed, barb
preserving and $\Ppar$-contextual relation.
The key to achieving this is to study the $\Ppar$-context closure of bisimilarity.
If we can demonstrate that this is reduction-closed then we have our
result.
To do this we must establish a decomposition theorem for interactions. For
instance, if $\pp$ and $\pq$ are bisimilar and we compose each of them
with a process $\pr$ then suppose
$$ \pp \Ppar \pr \redn S $$
represents an interaction between $\pp$ and $\pr$. We
decompose this into complementary actions 
$$\pp \lt{\lab} \pp' \quad \mbox{and} \quad \pr \lt{\dual{\lab}} \pr'$$
respectively. Note however that $S$ is not necessarily obtained 
by a parallel composition of the targets of the transitions: 
$\pp' \Ppar \pr'$. Instead, $\pp'$ and $\pr'$ may contain indirect references
and their corresponding resources. These need to be matched up
correctly to obtain $S$. We achieve this by introducing the 
%Consider two nodes of the labelled transition system
%$$
%(\cenv \sep \kenv \types \conf) \qquad \mbox{and} \qquad (\cenv \sep
%\kenv \types \confd)
%$$
%which are somehow \emph{complementary}. By this we mean that every
%indirect reference $\call{\tnk}$ in $\conf$ has a corresponding stored
%abstraction $\res{\tnk}{\valv}$ in $\confd$ and vice-versa. We
%
%introduce the 
\emph{merge} (partial) operator $\mg{\cdot}$ which will
match up these terms and replace every indirect reference to an
abstraction with the abstraction itself. 
%Of course these abstractions
%may contain indirect references themselves, so the order in which  
%the indirect references are merged is important. This is the reason
%$\kenv$ is maintained as an ordered set and why it becomes a parameter
%to the merge operation. We use the notation 
%$$
%\kenv \sts \tnk < \tnl
%$$
%to mean that $\tnk$ and $\tnl$ are members of $\kenv$ and appear in
%the order of $\kenv$ as $\tnk$ strictly before $\tnl$. 
We write
$$
\conf[\valv / \call{\tnk}]
$$
to denote the substitution of the value $\valv$ for every instance of
the indirect reference $\call{\tnk}$.
We define $\mg{\conf}$ then as the operator on terms
of the augmented syntax (up to $\equiv$) such that
  \begin{axioms}
     \mg{\conf}  & = & \conf \quad &  
    \mbox{if } \conf \mbox{ doesn't contain } \res{\tnk}{\valv} \mbox{
      for any } \tnk, \valv \\
%\mbox{ implies } \tnk \not\in \kenv 
    \mg{\newnt{\vec\ca}{\vec\typet}{\res{\tnk}{\valv} \Ppar
        \conf}} & = &                           
                          \mg{\newnt{\vec\ca}{\vec\typet}{\conf[\valv / \call{\tnk}]}} & 
    \mbox{if } \call{\tnk} \not\in \valv \\
  \end{axioms}
Intuitively, this says that we substitute any values stored at a
$\res{\tnk}{\valv}$ through for the corresponding $\call{\tnk}$. 
Note that this need not substitute for all
the indirect reference identifiers in $\conf$.
It is clear that the above definitions are only partial. For example, if
$\conf$ contains an occurrence of 
%$\call{\tnk}$ for
%which there is no corresponding 
$\res{\tnk}{\valv}$ for which $\call{\tnk}$ occurs in $\valv$,
then $\mg{\conf}$ is undefined.
In order to identify for which terms the merge is defined we make use
of the notion of \emph{reference graph}:
For a term $\conf$ we define the graph
$\rg{\conf}$ to be the graph which has nodes as the indirect reference
identifiers $\tnk$ in
$\conf$ and edges 
$$
\tnk \mapsto \tnl \quad \mbox{if} \quad \call{\tnl}\in \valv \quad
\mbox{for} \quad \res{\tnk}{\valv} \quad \mbox{in} \quad \conf
$$

\begin{proposition}
\label{prop:mg-ok}
%  If $\cenv \sep \kenv \types \conf$ and $\cenv \sep \kenv \types
%  \confd$ and moreover, $\kenv \sts \conf \merge \confd$ is defined,
%then $\cenv \types \conf \merge \confd$ is a valid judgement of
%\hopi.
$\mg{\cdot}$ is a well-defined partial function such that
$\mg{\conf}$
is defined if and only if $\rg{\conf}$ is acyclic.
\end{proposition}
\proof%\begin{pf}
  Given in Appendix~\ref{app:mg-ok}.
\qed%\end{pf}

\begin{lemma}[Composition/Decomposition] 
For $\cenv \sep \kenv \types \conf, \confd$ 
\begin{itemize}
\item [(i)] If $\mg{\conf \Ppar \confd} \equiv \confe$ and 
$$
(\cenv \sep \kenv \types \conf) \lt{\lab} (\cenv, \cenv' \sep \kenv,
\kenv' \types \conf')
\qquad
\mbox{and}
\qquad
(\cenv \sep \kenv \types \confd) \lt{\dual{\lab}} (\cenv, \cenv' \sep \kenv,
\kenv' \types \confd')
$$ 
then there exists a $\confe'$ such that $\confe \Redn \confe'$ and 
$
\mg{\newn{\cenv'}{\conf' \Ppar  \confd'}} = \confe'
$
\item [(ii)] If $\mg{\conf} \equiv E$ and $\conf \redn \conf'$ then
  there exists a $\confe'$ such that $\confe \redn \confe'$ and
  $\mg{\conf'} \equiv \confe'$
\item [(iii)] If $\mg{\conf \Ppar \confd} \equiv
  \confe$ and $\confe \redn \confe'$ then one of the following hold
  \begin{itemize}
  \item []$\conf \redn \conf'$ with  $\mg{\conf' \Ppar \confd} \equiv
\confe'$ 
\item [or]$\confd \redn \confd'$ with $\mg{\conf \Ppar
  \confd'} \equiv \confe'$ 
\item [or] $
(\cenv \sep \kenv \types \conf) \wlt{\lab} (\cenv, \cenv' \sep \kenv,
\kenv' \types \conf')
$
and
$
(\cenv \sep \kenv \types \confd) \wlt{\dual{\lab}} (\cenv, \cenv' \sep \kenv,
\kenv' \types \confd')
$
with  $\mg{\newn{\cenv'}{\conf' \Ppar \confd'}} \equiv \confe'$.
  \end{itemize}
  \end{itemize}
\end{lemma}
\proof%\begin{pf}
Part (ii) is straightforward as the merge operator $\mg{\;}$
simply removes subterm of the form $\res{\tnk}{\valv}$, which can't be
involved in reductions, and substitutes higher-order values through
for variables of higher-order type. Reductions are based on structure
alone except for the conditionals which can be affected by first-order
substitutions of channel names only.

To show (i) we must consider all the possible cases for $\lab$. By
symmetry there are four distinct pairs of complementary actions. We
only consider the cases where $\lab$ is $\newinlab{\ca}{\tnk}$ and
$\newinlab{\tnk}{\tnl}$ as the first-order actions can be treated similarly.
\begin{itemize}
\item [\textbf{Case:}] $\cenv \sep \kenv \types \conf
  \lt{\newinlab{\ca}{\tnk}}
\cenv \sep \kenv, \tnk : \ttrig{\typeu} \types \conf'$
and 
$\cenv \sep \kenv \types \confd \lt{\newoutlab{\ca}{\tnk}} \cenv \sep
\kenv , \tnk : \ttrig{\typeu} \types \confd'$.
By inspection we see that 
\begin{itemize}
\item $\conf \equiv \newn{\cenv'}{\inp{\ca}{\vx:\typet}{\pp} \Ppar
    \conf''} \qquad \mbox{with } \typet \tiso \tabs{\typeu}$
\item $\conf' \equiv \newn{\cenv'}{\app{\abst{\vx:\typet}{\pp}}{\call{\tnk}} \Ppar
    \conf''}$
\item $\confd \equiv \newn{\cenv''}{\out{\ca}{\valv}{\pq} \Ppar
    \confd''}$
\item $\confd' \equiv \newn{\cenv''}{\res{\tnk}{\valv} \Ppar \pq \Ppar \confd''}$ 
\end{itemize}
It is easy to see that $\mg{\conf \Ppar \confd} \redn 
\mg{\newn{\cenv', \cenv''}{\app{\abst{\vx:\typet}{\pp}}{\valv} \Ppar \conf'' \Ppar \pq
    \Ppar \confd'' }}$ let us call the target
  of this reduction $\confe'$.
We simply need to check 
$$
\begin{array}{llcl}
& \confe' & \equiv & \mg{\newn{\cenv', \cenv''}{\app{\abst{\vx:\typet}{\pp}}{\valv} \Ppar \conf'' \Ppar \pq
    \Ppar \confd'' }} \\
 (\call{\tnk}\not\in\valv) &      & \equiv & \mg{\newn{\cenv'}{\app{\abst{\vx:\typet}{\pp}}{\call{\tnk}} \Ppar
            \conf''} \Ppar \newn{\cenv''}{\res{\tnk}{\valv} \Ppar \pq
            \Ppar \confd''}} \\
      &  & \equiv & \mg{\conf' \Ppar \confd'}
\end{array}
$$
\item [\textbf{Case:}]  $\cenv \sep \kenv \types \conf
  \lt{\newinlab{\tnk}{\tnl}}
\cenv \sep \kenv, \tnl : \ttrig{\typet} \types \conf'$
and 
$\cenv \sep \kenv \types \confd \lt{\newoutlab{\tnk}{\tnl}} \cenv \sep
\kenv , \tnl : \ttrig{\typet} \types \confd'$.
Again, by inspection we see that 
\begin{itemize}
\item $\conf \equiv \newn{\cenv'}{\res{\tnk}{\valv} \Ppar
    \conf''}$
\item $\conf' \equiv \newn{\cenv'}{\app{\valv}{\call{\tnl}} \Ppar
    \res{\tnk}{\valv} \Ppar
    \conf''}$
\item $\confd \equiv \newn{\cenv''}{\app{\call{\tnk}}{\valw} \Ppar
    \confd''}$
\item $\confd' \equiv \newn{\cenv''}{\res{\tnl}{\valw} \Ppar \confd''}$ 
\end{itemize}
Note that the previous proposition tells us that 
$\rg{\conf \Ppar \confd}$ must be acyclic --- in
particular, $\call{\tnk} \not\in \valv$.
Here we see that 
$$
\begin{array}{llcl}
 & \mg{\conf \Ppar \confd} & \equiv & 
    \mg{\newn{\cenv', \cenv''}{\res{\tnk}{\valv} \Ppar
    \conf'' \Ppar \app{\call{\tnk}}{\valw} \Ppar
    \confd''}} \\
 (\call{\tnk}\not\in \valv) &  & \equiv & 
    \mg{\newn{\cenv', \cenv''}{\res{\tnk}{\valv} \Ppar
    \conf'' \Ppar \app{\valv}{\valw} \Ppar
    \confd''}} \\
 (\call{\tnl}\not\in \valv, \valw, \conf'', \confd'') &  & \equiv &
    \mg{\newn{\cenv', \cenv''}{\res{\tnk}{\valv} \Ppar
    \conf'' \Ppar \app{\valv}{\call{\tnl}} \Ppar \res{\tnl}{\valw} \Ppar
    \confd''}} \\
 & & \equiv & \mg{\conf' \Ppar \confd'}
\end{array}
$$
So by letting $\confe'$ be $\mg{\conf' \Ppar \confd'}$ 
we note that $\mg{\conf \Ppar \confd} \Redn \confe'$ as required.
\end{itemize}

\bigskip
To show (iii) we suppose $\mg{\conf \Ppar \confd} \equiv \confe$ and
that $\confe \redn \confe'$. We must consider all possible ways in
which this reduction can occur. If the reduction arises from a
conditional then it is clear that we must have $\conf \redn \conf'$ or
$\confd \redn \confd'$ for some $\conf'$ or $\confd'$. Moreover it is
easy to check that $\mg{\conf' \Ppar \confd}$ (resp $\mg{\conf \Ppar
  \confd'}$) $\equiv \confe'$.
There are two more possibilities to consider:
\begin{itemize}
\item [\textbf{Case:}] the reduction arises from a $\beta$-reduction.
In this case either $\conf \redn \conf'$ or $\confd \redn \confd'$ as
above and the result follows easily, or $\valv$ is $\abst{\vx:\typeu}{\pp}$ 
and
\begin{itemize}
\item $\conf \equiv \newn{\cenv'}{\app{\call{\tnk}}{\valw} \Ppar
    \conf''} \qquad \mbox{with all names in } \cenv' \mbox{ appearing
    in } \valw$
\item $\confd \equiv \newn{\cenv''}{\res{\tnk}{\valv}
    \Ppar \confd''} \qquad \mbox{with} \qquad \call{\tnk}\not\in\valv$
\item $\confe' \equiv \mg{\newn{\cenv', \cenv''}{
\pp [\valw / \vx] \Ppar \conf'' \Ppar \res{\tnk}{\valv}
    \Ppar \confd''}}$
\end{itemize}
or a symmetric version of these with the roles of $\conf$ and $\confd$
reversed.
So we notice that if  $\typeu \tiso \tabs{\typet}$, 
we have
$$
\cenv \sep \kenv \types \conf \lt{\newoutlab{\tnk}{\tnl}} \cenv \sep
\kenv, \tnl:\ttrig{\typet} \types \conf'
\qquad 
\mbox{and}
\qquad
\cenv \sep \kenv \types \confd \wlt{\newinlab{\tnk}{\tnl}} \cenv \sep
\kenv, \tnl:\ttrig{\typet} \types \confd'
$$
where $\conf' \equiv \newn{\cenv'}{\res{\tnl}{\valw} \Ppar \conf''}$
and 
$\confd' \equiv \newn{\cenv''}{\pp [\call{\tnl} /\vx] \Ppar
  \res{\tnk}{\valv} \Ppar \confd''}$.
We check:
$$
\begin{array}{llcl}
 & \mg{\conf' \Ppar \confd'} & \equiv &
 \mg{\newn{\cenv'}{\res{\tnl}{\valw} \Ppar \conf''} \Ppar 
 \newn{\cenv''}{\pp[ \call{\tnk} /\vx]} \Ppar
  \res{\tnk}{\valv} \Ppar \confd''} \\
 (\call{\tnl} \not\in \valv, \valw,\conf'' , \confd'') & & \equiv &
  \mg{\newn{\cenv', \cenv''}{ \conf'' \Ppar \pp [\valw / \vx] \Ppar
      \res{\tnk}{\valv} \Ppar \confd''}} \\
 & & \equiv & \confe'
\end{array}
$$
as required. Alternatively, it could be that $\typeu$ is a base
type, in which case
$$
\cenv \sep \kenv \types \conf
\lt{\newlab{\cenv'}{\outlab{\tnk}{\valw}}} \cenv, \cenv' \sep \kenv
\types \conf'
\qquad
\mbox{and}
\qquad
\cenv \sep \kenv \types \confd
\wlt{\newlab{\cenv'}{\inlab{\tnk}{\valw}}} \cenv, \cenv' \sep \kenv
\types \confd'
$$
where $\conf' \equiv \conf''$ and $\confd' \equiv
\newn{\cenv''}{\pp [\valw / \vx] \Ppar \res{\tnk}{\valv}
  \Ppar \confd''}$. It is easy to check that 
$\mg{\conf' \Ppar \confd'} \equiv \confe'$ as required.
\item [\textbf{Case:}] the reduction arises from communication. Again
  we see that either $\conf \redn \conf'$ or $\confd \redn \confd'$,
  in which case we easily obtain the result, or 
  \begin{itemize}
  \item $\conf \equiv \newn{\cenv'}{\out{\ca}{\valv}{\pp} \Ppar
      \conf''}$  
  \item $\confd \equiv \newn{\cenv''}{\inp{\ca}{\vx:\typet}{\pq} \Ppar
      \confd''}$
  \item $\confe' \equiv \mg{ \newn{\cenv'}{\pp \Ppar \conf''} \Ppar
      \newn{\cenv''}{\app{\abst{\vx:\typet}{\pq}}{\valv} \Ppar \confd''}}$
  \end{itemize}
or a symmetric version of this with the roles of $\conf$ and $\confd$
reversed.
Again we must consider whether the type $\typet$ is a base type or
higher-order. We omit the details of the former case. Suppose then
that $\cenv \sep \kenv \types \valv : \typet \tiso \tabs{\typeu}$
we know
$$
\cenv \sep \kenv \types \conf 
\lt{\newoutlab{\ca}{\tnk}}
\cenv \sep \kenv, \tnk:\typeu \types \conf'
\qquad 
\mbox{and}
\qquad
\cenv \sep \kenv \types \confd
\lt{\newinlab{\ca}{\tnk}}
\cenv \sep \kenv, \tnk:\typeu \types \confd'
$$
where
$\conf' \equiv \newn{\cenv'}{\res{\tnk}{\valv} \Ppar \pp \Ppar
  \conf''}$
and 
$\confd' \equiv \newn{\cenv''}{\app{\abst{\vx:\typet}{\pq}}{\call{\tnk}} \Ppar \confd''}$.
We check:
$$
\begin{array}{llcl}
 & \mg{\conf' \Ppar \confd'} & \equiv &
   \mg{\newn{\cenv'}{\res{\tnk}{\valv} \Ppar \pp \Ppar
  \conf''} \Ppar \newn{\cenv''}{\app{\abst{\vx:\typet}{\pq}}{\call{\tnk}} \Ppar \confd''}}
\\
 (\call{\tnk} \not\in \valv , \pp , \conf'' , \confd'') & & \equiv &
   \mg{\newn{\cenv', \cenv''}{\pp \Ppar \conf'' \Ppar \app{\abst{\vx:\typet}{\pq}}{\valv}
       \Ppar \confd''}} \\
& & \equiv & \confe'
\end{array}
$$
as required.\qed
\end{itemize}

\begin{definition}
Let $\mgctxtrel$ be defined to be 
$$
\cenv \sep \kenv \sts \mg{\conf_1 \Ppar \confd} \mgctxtrel
\mg{\conf_2 \Ppar \confd}
\qquad\mbox{if and only if}\qquad
\cenv \sep \kenv \sts \conf_1 \bisim \conf_2 \quad \mbox{and} \quad
\cenv \sep \kenv \types \confd
$$
whenever $\mg{\conf_1 \Ppar \confd}$ and $\mg{\conf_2 \Ppar \confd}$ are
defined.
\end{definition}
Note that in the case where $\kenv$ is empty we have that
$\mg{\conf_i \Ppar \confd}=\conf_i \Ppar \confd$, and hence
$\mgctxtrel$ and $\parctxteq$ coincide.
\begin{lemma} \label{lemma:mergecong}
  $\mgctxtrel$ is reduction-closed.
\end{lemma}
\proof%\begin{pf}
Follows easily from the previous lemma. Take 
$\cenv \sep \kenv \sts \mg{\conf_1 \Ppar  \confd} \mgctxtrel \mg{\conf_2 \Ppar
  \confd}$
and suppose 
$\mg{\conf_1 \Ppar \confd} \redn 
\confe$. We must show that 
$\mg{\conf_2 \Ppar \confd} \redn \confe'$
for some $\confe'$ such that $\cenv \sep \kenv \sts \confe \mgctxtrel
\confe'$.
We know from Part (iii) of the previous lemma that one of three cases
must hold. Either, $\conf_1 \redn \conf_1'$, $\confd \redn \confd'$ or
there are complementary actions from both $\conf_1$ and $\confd$. We
only deal with the last case as the others follow easily from the
hypothesis that $\cenv \sep \kenv \sts \conf_1 \bisim \conf_2$ and
Part (ii) of the previous lemma.

We have then that 
$\cenv \sep \kenv \types \conf_1 \wlt{\lab} \cenv,
\cenv' \sep \kenv, \kenv' \types \conf_1'$ 
and 
$\cenv \sep \kenv \types \confd \wlt{\dual{\lab}} \cenv, \cenv' \sep \kenv, \kenv'
\types \confd'$ 
such that 
$\confe \equiv \mg{\conf_1' \Ppar  \confd'}$. 
We know by hypothesis that there must exist some
$$
\cenv \sep \kenv \types \conf_2 \wlt{\lab} \cenv, \cenv' \sep \kenv,
\kenv' \types \conf_2' 
$$
such that 
$$\cenv, \cenv' \sep \kenv, \kenv' \sts \conf_1' \bisim
\conf_2'. \qquad (\dagger)$$
We can now use Parts (i) and (ii) of the previous lemma to see that 
$\mg{\conf_2 \Ppar \confd} \Redn \confe'$ such that 
$\confe' \equiv \mg{\conf_2' \Ppar \confd'}$. 
Note that $(\dagger)$ guarantees 
$\cenv \sep \kenv \sts \confe \mgctxtrel \confe'$ to finish.
\qed%\end{pf}

\begin{theorem} For all closed terms $\pp, \pq$ of \hopi:
  $$ \cenv \sts \pp \bisim \pq \quad \mbox{implies} \quad \cenv \sts
  \pp \parctxteq \pq$$
\end{theorem}
\proof%\begin{pf}
We let $\parbisim$ denote the relation 
$$
\cenv, \cenv' \sts (\pp \Ppar \pr) \parbisim (\pq \Ppar \pr) \mbox{ iff } 
\cenv \sts \pp \bisim \pq \mbox{ and } \cenv, \cenv' \types \pr  
$$
It is easy to see that $\parbisim$ is a 
$\Ppar$-contextual relation over terms of \hopi.
It is also easy to see that $\parbisim$ is symmetric and barb
preserving and coincides with $\mgctxtrel$ for closed terms of \hopi, thus
Lemma~\ref{lemma:mergecong} can be instantiated to 
demonstrate that $\parbisim$ is reduction-closed and,
given that $\parctxteq$ is defined to be the
largest symmetric, $\Ppar$-contextual, reduction-closed, and barb-preserving
relation over terms of \hopi, then we have our result.  
\qed%\end{pf}

\begin{corollary}[Soundness] \label{cor:soundness}
For all terms $\pp, \pq$ of \hopi:
$$
\tenv \sts \pp \openex{\bisim} \pq \quad \mbox{implies} \quad \tenv \sts
\pp \ctxteq \pq
$$
\end{corollary}
\proof%\begin{pf}
Follows from the previous theorem and Lemma~\ref{lemma:context-lemma}.
\qed%\end{pf}

%}}}
%{{{ Completeness

\subsection{Completeness of bisimilarity for contextual equivalence}

The interactions described by the labelled transition system 
are not obviously derived by genuine contextual observations in 
\hopi~because of the use of the extra syntax for indirect references. 
In order to show completeness of our bisimilarity for contextual
equivalence we must demonstrate that the indirect references are in
fact definable as terms of the language proper.
Following Sangiorgi \cite{SangiorgiBook}, we implement the implicit
protocol outlined by the indirect references by using the following
translation of the augmented terms into \hopi:
\begin{axioms}
\sem{\tnk_1: \ttrig{\typet_1}, \ldots, \tnk_n : \ttrig{\typet_n}} & =
& \tnk_1 : \tchan{\typet_1}, \ldots , \tnk_n : \tchan{\typet_n} \\
\sem{\tenv \sep \kenv \types \conf} & = & \tenv , \sem{\kenv} \types
\sem{\conf}_{\kenv} \\
\sem{\call{\tnk}}_{\kenv}  & = &
\abst{\vx:\typet}{\out{\tnk}{\vx}{\nil}} \quad & \mbox{if }
\kenv(\tnk) = \ttrig{\typet} \\
\sem{\res{\tnk}{\valv}}_{\kenv} & = &
\repl{\tnk \sem{\valv}_\kenv }  & 
\end{axioms}
The translation acts homomorphically on all other terms. 
We abuse notation here by using identifiers $\tnk$ as channel names in
the translation.
It is evident
that this translation is well-defined in the sense that the
translation of well-typed augmented terms are indeed well-typed terms
of \hopi.

We would now like to prove a correspondence between reductions from the
terms of the augmented syntax and reductions between their
translations.
However, we note that in translating a  term containing both 
$\res{\tnk}{\valv}$ and $\call{\tnk}$ we provide matching input 
and output prefixes, which, in \hopi~may create a communication 
which was not possible in the source term. This turns out not to be of
particular concern to us though as we see that if we starting with
terms of \hopi, then terms reachable by transitions are
\emph{balanced} in the following sense:
we call a term $\conf$ of the augmented language
\emph{balanced} if for each $\tnk$ then $\conf$ contains at most one
of $\call{\tnk}$ (possible multiple times) or $\res{\tnk}{\valv}$.
Unfortunately the translation may introduce extra
reductions which aren't present in the source term. 
These arise through the translation of terms of the form
$\app{\call{\tnk}}{\valv}$. Note that 
$$\sem{\app{\call{\tnk}}{\valv}} =
\app{\abst{\vx:\typet}{\out{\tnk}{\vx}{\nil}}}{\sem{\valv}} \lt{\taulab}
\out{\tnk}{\sem{\valv}}{\nil} $$
but $\app{\call{\tnk}}{\valv}$ has no corresponding reduction. 
We will identify these rogue reductions as housekeeping reductions and
indicate them with $\lt{\hbeta}$ defined as any reduction
which can be derived using the axiom  
$$
(\hbeta-\mbox{redn}) \qquad \app{\abst{\vx:\typet}{\out{\tnk}{\vx}{\nil}}}{\valv} \lt{} \out{\tnk}{\valv}{\nil}
$$
\begin{lemma} \label{lemma:housebeta} If $\cenv \sep \kenv \types \conf$ is balanced then 
  \begin{enumerate}
  \item \label{part:housebeta-implies}
    If $\conf \Redn \conf'$ then $\sem{\conf}_{\kenv} \Redn
    \sem{\conf'}_{\kenv}$
  \item \label{part:housebeta-implied}
   If $\sem{\conf}_{\kenv} \Redn \pp$ then 
   $\sem{\conf}_{\kenv} \Redn \sem{\confd}_{\kenv} \transhbeta \pp$
for some
    $\cenv \sep \kenv \types \confd$ such that $\conf \Redn \confd$.
  \end{enumerate}
\end{lemma}

\iflong

\proof%\begin{pf}
We will omit mention of the environment $\kenv$ in the proof as it
plays no role.
Part \ref{part:housebeta-implies} is straightforward. For Part \ref{part:housebeta-implied} we 
use induction on the length of the reductions.
If there are no reductions then we are done.
We examine the base case
in which $\sem{\conf} \redn \pp$.
If this reduction happens to be a housekeeping move, that is,
$\sem{\conf} \lt{\hbeta} \pp$ then there is nothing to prove.
Suppose otherwise, then it is not too difficult to check that 
$\pp \equiv \sem{\confd}$ for some $\confd$ such that
$\conf \redn \confd$.
For the inductive case suppose that
$$\sem{\conf} \redn \Redn \pp \qquad (\dagger)$$
By inspecting the translation $\sem{\cdot}$ and using the fact that
$\conf$ is balanced 
we see that 
$$\sem{\conf} \lt{\hbeta} \redn \pq \qquad \mbox{implies} \qquad  \sem{\conf}
\redn \lt{\hbeta} \pq$$
thus we may assume that the first reduction in $(\dagger)$ above is
not of the form $\lt{\hbeta}$. This means that 
$\sem{\conf} \redn \sem{\conf'} \Redn \pp$ 
for some $\conf'$ such that $\conf \redn \conf'$. It is clear that
$\conf'$ is also balanced so we may apply the inductive hypothesis to 
$$\sem{\conf'} \Redn \pp$$ to obtain a $\confd$ such that 
$\conf' \Redn \confd'$ and $\sem{\conf'} \Redn \sem{\confd} \transhbeta \pp$.
Putting these together we obtain 
$$
\conf \redn \conf' \Redn \confd \qquad \mbox{and} \qquad \sem{\conf}
\redn \sem{\conf'} \Redn \sem{\confd} \transhbeta \pp
$$
as required.
\qed%\end{pf}
\fi
When $\cenv'$ is of length at most one, we shall write
$\succsucc{\cenv'}$ as shorthand, defined:
\[
  \succsucc{\emptyset}=\succsucc{\unit} \quad
  \succsucc{\ca:\typet}=\succsucc{\ca}
\]
Moreover, note that whenever
$(\cenv \sep \kenv \types \confd) \wlt{\lab} (\cenv, \cenv' \sep \kenv,
\kenv' \types \confd')$, we have that $\cenv'$ has at length most one,
and so $\succsucc{\cenv'}$ is well-defined.

\begin{proposition} \label{prop:contexts-labels}
  For each $\lab, \cenv$ and fresh channels $\succbarb, \failbarb$ of appropriate
  type given by $\lab$ and $\cenv$,
  there exists a process $\ctxtlab{\cenv}{\lab}$ 
  (defined in Figure~\ref{fig:contexts-for-labels})
  in \hopi~such that 
 if 
$$\cenv \sep \kenv \types \conf \lt{\lab} \cenv, \cenv' \sep \kenv,
 \kenv' \types \conf'$$
 then 
$$\cenv , \sem{\kenv, \kenv'},
 \succbarb:\tchan{\typet_0}, \failbarb : \tchan{\tunit} \types
 \ctxtlab{\cenv, \sem{\kenv}}{\lab}$$ 
and moreover, for balanced $\confd$
$$
(\cenv \sep \kenv \types \confd) \wlt{\lab} (\cenv, \cenv' \sep \kenv,
\kenv' \types \confd') 
$$
if and only if $\cenv \sep \kenv \types \confd$ and 
$$
%(\cenv, \sem{\kenv,
%  \kenv'} , \succbarb: \tchan{\typet}, \failbarb: \tchan{\tunit}
%\types 
\ctxtlab{\cenv, \sem{\kenv}}{\lab} \Ppar
\sem{\confd}_{\kenv}
%)
\Redn 
%(\cenv, \sem{\kenv, \kenv'} , \succbarb :
%\tchan{\typet}, \failbarb: \tchan{\tunit} \types 
\newn{\cenv'}{\succsucc{\cenv'} \Ppar \pp}
%)
\qquad \mbox{with} \qquad 
\sem{\confd'}_{\kenv, \kenv'} \transhbeta \pp.$$
\end{proposition}

\iflong

\proof%\begin{pf}
It is straightforward to check that $\cenv, \sem{\kenv, \kenv'},
\succbarb:\tchan{\typet_0}, \failbarb:\tchan{\tunit} \types
\ctxtlab{\cenv}{\lab}$ whenever 
$$\cenv \sep \kenv \types \conf \lt{\lab} \cenv, \cenv' \sep \kenv, \kenv' \types \conf'.$$
For the remainder, to show the `only if' direction we use
Lemma~\ref{lemma:housebeta} Part~\ref{part:housebeta-implies} 
to reduce our obligation to the case of a single transition $\lt{\lab}$, 
and we must consider each label $\lab$. By
way of example we show the case for $\lab =
\newoutlab{\tnk}{\tnl}$ (the other cases can be treated similarly).
Suppose:
$$
(\cenv \sep \kenv \types \confd) \lt{\lab} (\cenv \sep \kenv,
\tnl:\ttrig{\typeu} \types \confd'). 
$$
then we know that 
$$\confd \equiv \newn{\cenv''}{\app{\call{\tnk}}{\valv} \Ppar
  \confd''}$$
and
$$ \confd' \equiv \newn{\cenv''}{\res{\tnl}{\valv} \Ppar \confd''}.$$
We see that for $\typet \tiso \tabs{\typeu}$
$$
\begin{array}{lcl}
\ctxtlab{\cenv, \sem{\kenv}}{\lab} \Ppar \sem{\confd}_{\kenv} & \equiv & 
  \inp{\tnk}{\vx:\typet}{( \repl{
        \inp{\tnl}{\vy:\typeu}{\app{\vx}{\vy}} } \Ppar \succout{} ) }
    \Ppar
    \newn{\cenv''}{\app{(\abst{\vz:\typet}{\out{\tnk}{\vz}{\nil}})}{\sem{\valv}_{\kenv}} \Ppar \sem{\confd''}_{\kenv}} \\
 & \Redn  & \succout{} \Ppar \newn{\cenv''}{ \repl{
     \inp{\tnl}{\vy:\typeu}{\app{\sem{\valv}_{\kenv}}{\vy}} \Ppar
     \sem{\confd''}_{\kenv}} } \\
 & \Redn & \succsucc{} \Ppar \sem{\confd'}_{\kenv, \tnl:\ttrig{\typeu}}   
\end{array}
$$
as required.

For the converse direction we suppose that 
$$
\ctxtlab{\cenv, \sem{\kenv}}{\lab} \Ppar \sem{\confd}_{\kenv} \wlt{}
\newn{\cenv'}{\succsucc{\cenv'} \Ppar \pp}
$$
Again, we must perform a case analysis on $\lab$. We show the case in
which $\lab$ is $\newinlab{\tnk}{\tnl}$
 (the other cases can be treated similarly).
We know $\cenv'$ is empty so
$\ctxtlab{\cenv, \sem{\kenv}}{\lab} \Ppar \sem{\confd}_{\kenv} \wlt{}
\succsucc{} \Ppar \pp$.
Note that $\ctxtlab{\cenv, \sem{\kenv}}{\lab}$ has no reductions of its
own and can only interact with $\sem{\confd}_{\kenv}$ so 
we can detail the assumed reductions as 
$$
\ctxtlab{\cenv, \sem{\kenv}}{\lab} \Ppar \sem{\confd}_{\kenv} \wlt{} 
\ctxtlab{\cenv, \sem{\kenv}}{\lab} \Ppar \pp_0 \lt{} 
\succout{} \Ppar \pp_1 \wlt{} \succsucc{} \Ppar \pp
$$
where $\sem{\confd} \wlt{} \pp_0$ and $\pp_1 \wlt{} \pp$.
We assumed that $\confd$ is balanced so Lemma~\ref{lemma:housebeta}
Part \ref{part:housebeta-implied}
applied to $\sem{\confd} \Redn \pp_0$ tells us that 
$\sem{\confd} \Redn \sem{\confd_0}_{\kenv} \transhbeta \pp_0$ 
for some $\confd_0$ such that 
$\confd \wlt{} \confd_0$. 
We know that $\pp_0$ is obtained from $\sem{\confd_0}_{\kenv}$ by housekeeping
reductions and that it interacts with $\ctxtlab{\cenv}{\lab}$. This
tells us that we must have the forms 
$$
\pp_0 \equiv \newn{\cenv''}{\repl{ \tnk \sem{\valv}_{\kenv}} \Ppar \pp_0'}
$$
and
$$
\pp_1 \equiv
\newn{\cenv''}{\app{\sem{\valv}_{\kenv}}{\sem{\call{\tnl}}_{\kenv,
      \tnl : \ttrig{\typeu}}} \Ppar \repl{ \tnk \sem{\valv}_{\kenv} \Ppar \pp_0'}}
$$
This in turn tells us that 
$$
\confd_0 \equiv \newn{\cenv''}{\res{\tnk}{\valv} \Ppar \confd_0'}
$$
such that $\sem{\confd_0'}_{\kenv} \transhbeta \pp_0'$.
Now it is clear that 
$$
(\cenv \sep \kenv \types \confd_0) \lt{\newinlab{\tnk}{\tnl}}
(\cenv \sep \kenv, \tnl:\ttrig{\typeu}) \types \confd_1)
$$
where $\confd_1 \equiv \newn{\cenv''}{\app{\valv}{\call{\tnl}} \Ppar
  \res{\tnk}{\valv} \Ppar \confd_0'}$.
We check
$$
\begin{array}{lcl}
\sem{\confd_1}_{\kenv, \tnl:\ttrig{\typeu}} & \equiv & \newn{\cenv''}{
  \app{\sem{\valv}_{\kenv} }{\sem{\call{\tnl}}_{\kenv, \tnl :
      \ttrig{\typeu}}} \Ppar \repl{ \tnk \sem{\valv} } \Ppar
  \sem{\confd_0'}_{\kenv} } \\
 & \transhbeta & \newn{\cenv''}{ \app{\sem{\valv}_{\kenv} }{\sem{\call{\tnl}}_{\kenv, \tnl :
      \ttrig{\typeu}}} \Ppar \repl{ \tnk \sem{\valv} } \Ppar \pp_0' }
\\
 & \equiv & \pp_1 \\
 & \Redn & \pp
\end{array}
$$
Therefore $\sem{\confd_1} \Redn \pp$ and we can apply
Lemma~\ref{lemma:housebeta} Part~\ref{part:housebeta-implied} to this to see that 
$\sem{\confd_1} \Redn \sem{\confd'} \transhbeta \pp$ for some
$\confd'$ such that
$\confd_1 \Redn \confd'$.
By collecting the above together we obtain
$$
(\cenv \sep \kenv \types \confd) \Redn (\cenv \sep \kenv \types
\confd_0) \lt{\lab} (\cenv \sep \kenv, \tnl: \ttrig{\typeu} \types
\confd_1) \Redn (\cenv \sep \kenv, \tnl: \ttrig{\typeu} \types \confd')
$$
with $\sem{\confd'}_{\kenv, \tnl: \ttrig{\typeu}} \transhbeta \pp$ as required.
\qed%\end{pf}
\fi
%
%{{{ contexts for labels figure
%
\begin{figure}

$$
\begin{array}{lcll}
    \ctxtlab{\cenv}{\inlab{\dd}{\valv}} & = &
    \out{\dd}{\valv}{\succout{}}  & \\
    \ctxtlab{\cenv}{\outlab{\dd}{\valv}} & = & 
    \inp{\dd}{\vx:\typet}{\ifexp{\vx}{\valv}{\succout{}}{\nil}} 
     &
    \mbox{ where } \cenv(\dd) = \tchan{\typet} \\
    \ctxtlab{\cenv}{\newlab{\cb}{\inlab{\dd}{\cb}}} & = &
    \newn{\cb:\typet}{\out{\dd}{\cb}{\succout{\cb}}}  & 
    \mbox{ where } \cenv(\dd) = \tchan{\typet} \\
    \ctxtlab{\cenv}{\newlab{\cb}{\outlab{\dd}{\cb}}} & = &
    \inp{\dd}{\vx:\typet}{\ifoneexp{\vx\not\in\cenv}{\succout{\vx}}{\nil}}  & 
    \mbox{ where } \cenv(\dd) = \tchan{\typet} \\
    \ctxtlab{\cenv}{\newinlab{\dd}{\tnk}} & = & 
    \out{\dd}{\abst{\vx:\typeu}{\out{\tnk}{\vx}{\nil}}}{\succout{}} & 
    \mbox{ where } \cenv(\dd) = \tchan{\typet} \mbox{ and } \typet
    \tiso \tabs{\typeu} \\
    \ctxtlab{\cenv}{\newoutlab{\dd}{\tnk}} & = & 
    \inp{\dd}{\vx:\typet}{( \repl{
        \inp{\tnl}{\vy:\typeu}{\app{\vx}{\vy}} } \Ppar \succout{} ) }
     & 
      \mbox{ where } \cenv(\dd) = \tchan{\typet} \mbox{ and } \typet
    \tiso \tabs{\typeu} \\[\medskipamount]
    \multicolumn{4}{c}{\mbox{$\ic$ represents an encoding of internal choice in \hopi}}\\
    \multicolumn{4}{c}{\ifoneexp{\vx\not\in\emptyset}{\pp}{\pq} = \pp}\\
    \multicolumn{4}{c}{\ifoneexp{\vx\not\in(\ca:\typet,\cenv)}{\pp}{\pq} = 
        \ifoneexp{\vx=\ca}{\pq}{\ifoneexp{\vx\not\in\cenv}{\pp}{\pq}}}
\end{array}
$$

  \caption{Testing processes for labelled transitions}
    \label{fig:contexts-for-labels}

\end{figure}
%
%}}}

\begin{lemma}[Extrusion] \label{lemma:extrusion}
If $\cenv \sts \newn{\cenv'}{ \succsucc{\cenv'} \Ppar \pp } \parctxteq
\newn{\cenv'}{ \succsucc{\cenv'} \Ppar \pq }$
then $\cenv, \cenv' \sts \pp \parctxteq \pq$.
\end{lemma}

\iflong

\proof%\begin{pf}
  Follows a similar argument found in \cite{JeffreyRathke:tbcmlln}: define a
  relation $\rel$ such that
$$
\cenv , \cenv' \sts \pp \rel \pq 
\qquad 
\mbox{iff} 
\qquad 
\cenv \sts \newn{\cenv'}{ \succsucc{\cenv'} \Ppar \pp } \parctxteq
\newn{\cenv'}{ \succsucc{\cenv'} \Ppar \pq }
$$
and show that $\rel$ is barb-preserving, reduction-closed and
$\Ppar$-contextual.
These properties follow from the corresponding property for $\parctxteq$ 
and an extra piece of context to interact with $\succsucc{\cenv'}$.
\qed%\end{pf}

\fi

\begin{theorem}[Completeness]
For all closed terms $\pp, \pq$ of \hopi:
$$
\cenv \sts \pp \parctxteq \pq \qquad \mbox{implies} \qquad 
\cenv \sts \pp \bisim \pq
$$ 
\end{theorem}
\proof%\begin{pf}
  We define $\rel$ over terms of the augmented language  to be 
$$
\cenv \sep \kenv \sts \conf \rel \confd \qquad \mbox{iff} \qquad
\cenv, \sem{\kenv} \sts \sem{\conf}_{\kenv} 
\parctxteq \sem{\confd}_{\kenv}
$$
and show that $\rel$ is a bisimulation.
Take $\cenv \sep \kenv \sts \conf \rel \confd$ and suppose that
$$(\cenv \sep \kenv \types \conf) \lt{\lab} (\cenv, \cenv' \sep \kenv ,
\kenv' \types \conf').$$
We know from Proposition~\ref{prop:contexts-labels} that 
$$
\cenv, \sem{\kenv, \kenv'}, \succbarb:\tchan{\typet_0}, \failbarb :
\tchan{\tunit} \types \ctxtlab{\cenv, \sem{\kenv}}{\lab}
$$
and that
$$
\ctxtlab{\cenv, \sem{\kenv}}{\lab} \Ppar \sem{\conf}_{\kenv} \Redn 
 \newn{\cenv'}{ \succsucc{\cenv'} \Ppar \pp}
$$
with $\sem{\conf'}_{\kenv, \kenv'} \transhbeta \pp$.
We know that 
$$
\cenv, \sem{\kenv} \sts \sem{\conf}_{\kenv} 
\parctxteq \sem{\confd}_{\kenv}
$$
by the definition of $\rel$, and hence, by contextuality we also have
$$
\cenv, \sem{\kenv, \kenv'}, \succbarb :\tchan{\typet_0}, \failbarb :
\tchan{\tunit} \sts \ctxtlab{\cenv, \sem{\kenv}}{\lab} \Ppar \sem{\conf}_{\kenv} 
\parctxteq \ctxtlab{\cenv, \sem{\kenv}}{\lab} \Ppar \sem{\confd}_{\kenv}
$$
This tells us that
$$
\ctxtlab{\cenv, \sem{\kenv}}{\lab} \Ppar \sem{\confd}_{\kenv} \Redn \pq'
$$
such that 
$$\cenv , \sem{\kenv, \kenv'} \sts \newn{\cenv'}{ \succsucc{\cenv'}
  \Ppar \pp} \parctxteq \pq'. \qquad (\dagger)$$ 
But by the construction of $\ctxtlab{\cenv,\sem{\kenv}}{\lab}$ we
notice that $\newn{\cenv'}{ \succsucc{\cenv'} \Ppar \pp}$ barbs on
$\succbarb$ but not
on $\failbarb$. 
Therefore, by the preservation of barbs property of $\parctxteq$, we
know that $\pq'$ must also barb on $\succbarb$ but not on $\failbarb$.
This constrains $\pq'$ so that $\pq' \equiv \newn{\cenv'}{
  \succsucc{\cenv'} \Ppar \pq}$. We apply Lemma~\ref{lemma:housebeta}
Part \ref{part:housebeta-implied}
to $\ctxtlab{\cenv, \sem{\kenv}}{\lab} \Ppar \sem{\confd}_{\kenv}
\Redn \pq'$ to see that there is some $\confd''$ such that
$\ctxtlab{\cenv, \sem{\kenv}}{\lab} \Ppar \sem{\confd}_{\kenv} \Redn \sem{\confd''}_{\kenv,\kenv'} \transhbeta
\newn{\cenv'}{\succsucc{\cenv'} \Ppar \pq}$ from which it clearly
follows that 
$\confd'' \equiv \newn{\cenv'}{\succsucc{\cenv'} \Ppar \confd'}$
and $\sem{\confd'}_{\kenv, \kenv'} \transhbeta \pq$. 
We use Proposition~\ref{prop:contexts-labels} again to see that 
$$(\cenv \sep \kenv \types \confd) \wlt{\lab} (\cenv, \cenv' \sep
\kenv, \kenv' \types \confd')$$
and we now must show that $\cenv, \cenv' \sep \kenv, \kenv' \sts
\conf' \rel \confd'$.
To do this we use Lemma~\ref{lemma:extrusion} on $(\dagger)$ (note
that $\pq' \equiv \newn{\cenv'}{\succsucc{\cenv'} \Ppar \pq}$) 
to see that 
$\cenv, \cenv' , \sem{\kenv, \kenv'} \sts \pp \parctxteq \pq$.
It is also easy to check that $\hbeta$-reductions are confluent with
respect to all other reductions and hence
 preserve contextual
equivalence, that is $\transhbeta \subseteq \parctxteq$, so we also
have
$\cenv, \cenv' , \sem{\kenv, \kenv'} \sts \sem{\conf'}_{\kenv, \kenv'}
\parctxteq \sem{\confd'}_{\kenv, \kenv'}$
because $\sem{\conf'}_{\kenv, \kenv'} \transhbeta \pp$ and 
$ \sem{\confd'}_{\kenv, \kenv'} \transhbeta \pq$.
This allows us to conclude $\cenv, \cenv' \sep \kenv, \kenv' \sts
\conf' \rel \confd'$ as required.

We must also consider transitions of the form
$$
(\cenv \sep \kenv \types \conf) \lt{\taulab} (\cenv, \cenv' \sep \kenv ,
\kenv' \types \conf').
$$
These can be dealt with as above but in this case no
$\ctxtlab{\cenv}{\lab}$
is needed.
\qed%\end{pf}
\begin{corollary}[Full abstraction]
  For all terms $\pp, \pq$ of \hopi:
$$\tenv \sts \pp \openex{\bisim} \pq \qquad \mbox{if and only if}
\qquad \tenv \sts \pp \ctxteq \pq$$
\end{corollary}
\proof%\begin{pf}
  Follows from Corollary~\ref{cor:soundness},
  Lemma~\ref{lemma:context-lemma},
  and the previous theorem.
\qed%\end{pf}

%}}}
%{{{ Conclusions

\section{Concluding remarks}

We have re-examined the use of labelled transitions to characterise
contextual equivalence in the higher-order $\pi$ calculus. The
technique of augmenting the core syntax with extra operators to assist
in the definition of the labelled transitions allows use to give a
direct proof of soundness of bisimilarity for contextual
equivalence. This advances Sangiorgi's analagous result by allowing
recursive types also. 

We believe that the technique of using extra operators to
describe the \emph{points of interaction} with the environment
in the lts is fairly robust and should be applicable to many
higher-order languages. Indeed, this was the approach that the authors
developed for their work on concurrent objects \cite{JeffreyRathke:famtsco}.

We have only concerned ourselves with the characterisation of
contextual equivalence in \hopi~and so far have not studied
Sangiorgi's translation of higher-order to first-order mobility.
Thus, the restriction to finite types for his translation is still
necessary.  It would be interesting to investigate whether the current
work could be of use in removing this type restriction for his
translation also.

%}}}

\appendix

\section{Proof of The Context Lemma}
\label{app:context-lemma}

We recall the statement of Lemma~\ref{lemma:context-lemma} and 
detail its proof here.
\[ \tenv \sts \pp \ctxteq \pq \quad \mbox{if and only if} \quad 
\tenv \sts \pp \parctxteq \pq.\]
The force of this lemma is to show that the simplified form of
observational testing allowed by $\parctxteq$ is sufficient to capture
the power of full contextual testing.  In order to prove this we
essentially need to show that $\parctxteq$ is preserved by the
operators of \hopi.  For the most part, this can be done directly and
is stated in Lemma~\ref{lemma:cong} below. 
\begin{lemma} \mbox{}
\label{lemma:cong}
\begin{enumerate} 
\item If $\cenv, \vx:\typet \sts \pp \parctxteq \pq$ and $\cenv \types \valv : \typet$
then $\cenv \sts \app{\abst{\vx:\typet}{\pp}}{\valv} \parctxteq \app{\abst{\vx:\typet}{\pq}}{\valv}$.
\item \label{item:cong-input} If $\cenv, \vx:\typet \sts \pp \parctxteq \pq$ and $\cenv \types \ca:\tchan{\typet}$ then
$\cenv \sts \inp{\ca}{\vx:\typet}{\pp} \parctxteq \inp{\ca}{\vx:\typet}{\pq}$.
\item If $\cenv \sts \pp \parctxteq \pq$, $\cenv \types \valw : \typet$ and $\cenv \types \ca:\tchan{\typet}$
then
$\cenv \sts \out {\ca}{\valw}{\pp} \parctxteq \out{\ca}{\valw}{\pq}$.
\item   If $\cenv \sts \pp_1 \parctxteq \pq_1$ and $\cenv \sts \pp_2 \parctxteq \pq_2$ then 
$\cenv \sts \ifexp{\valv}{\valw}{\pp_1}{\pp_2} \parctxteq  \ifexp{\valv}{\valw}{\pq_1}{\pq_2}$.
\item \label{item:cong-nu} If $\cenv, \ca : \typet \sts \pp \parctxteq \pq$ then
$\cenv \sts \newnt{\ca}{\typet}{\pp} \parctxteq  \newnt{\ca}{\typet}{\pq}$.
\item If $\cenv \sts \pp_1 \parctxteq \pq_1$ and $\cenv \sts \pp_2 \parctxteq \pq_2$ then
$\cenv \sts \pp_1 \Ppar \pp_2 \parctxteq \pq_1 \Ppar \pq_2$.
\item If $\cenv \sts \pp \parctxteq \pq$ then $\cenv \sts \repl{\pp} \parctxteq \repl{\pq}$.  
\end{enumerate}
\end{lemma}
\proof%\begin{pf}
The majority of these are straightforward by exhibiting appropriate
symmetric, reduction-closed, $\Ppar$-contextual, barb-preserving relations.
As an example of this we show the case for input prefixing (Case~\ref{item:cong-input}). 
We define $\rel$ so that ${\parctxteq} \subseteq {\rel}$ and moreover
$$
\cenv \sts \inp{\ca}{\vx:\typet}{\pp} \Ppar \pr
\rel \inp{\ca}{\vx:\typet}{\pq} \Ppar \pr \mbox{ for any  $\cenv \types \pr$}  \qquad (\dagger)
$$
It is clear that $\rel$ is symmetric, barb-preserving and
$\Ppar$-contextual 
so if we can show that it is reduction-closed then we may conclude that
$\rel$ coincides with $\parctxteq$ and we have our result.

Suppose that $(\dagger)$ holds and 
$$\inp{\ca}{\vx:\typet}{\pp} \Ppar \pr \redn \pp'.$$
We know then that either $\pr \redn \pr'$ and 
$\pp' \equiv \inp{\ca}{\vx:\typet}{\pp} \Ppar \pr'$ 
or the reduction came about by interaction, that is
$\pr \equiv \newn{\cenv'}{\out{\ca}{\valv}{\pr''} \Ppar \pr'''}$ with
$\ca \not\in \cenv'$ and by writing $\pr'$ for $\pr'' \Ppar \pr'''$
we have
$\pp' \equiv \newn{\cenv'}{\pp[\valv/\vx] \Ppar \pr'}$ for some
$\cenv, \cenv' \types \valv$ and $\cenv, \cenv' \types \pr'$.
If the former is true then we see immediately that
$$
\inp{\ca}{\vx:\typet}{\pq} \Ppar \pr \redn \inp{\ca}{\vx:\typet}{\pq}
\Ppar \pr'
$$
where
$$
\cenv \sts \inp{\ca}{\vx:\typet}{\pp} \Ppar \pr'
\rel \inp{\ca}{\vx:\typet}{\pq} \Ppar \pr'.
$$
If instead the latter is true then we use the fact that 
$$
\cenv, \vx:\typet \sts \pp \parctxteq \pq 
$$
to see that 
$\cenv,  \cenv' \sts \pp[\valv/\vx] \parctxteq
\pq[\valv/\vx]$
and note that
$$\inp{\ca}{\vx:\typet}{\pq} \Ppar \pr \redn \newn{\cenv'}{\pq[\valv/\vx] \Ppar \pr'}$$
where (using $\Ppar$-contextuality and Case~\ref{item:cong-nu}) 
$$
\cenv \sts \newn{\cenv'}{\pp[\valv/\vx] \Ppar
  \pr'} \parctxteq \newn{\cenv'}{\pq[\valv/\vx] \Ppar \pr'}
$$
as required.
\qed%\end{pf}
Notice that there are two
particular cases which are not covered by this lemma: application of a
function to, and output of higher-order $\parctxteq$-related values
(c.f.~Corollary~\ref{corollary:cong-hard}).
Establishing that $\parctxteq$ is preserved in these cases can be done
directly but is a little more involved. We notice that the property we require in 
both cases follows immediately from \emph{Substitutivity} (cf.~Corollary~\ref{corollary:substitutivity})
, that is (ignoring types):
\[ \mbox{if } 
\pp \parctxteq \pq \mbox{ then } \pr [\abst{\vx}{\pp} /\vy ] \parctxteq 
 \pr [\abst{\vx}{\pq} /\vy].
\]
The remainder of the appendix is devoted to achieving this. The proof follows a 
very similar scheme to the proof of Proposition~4.2.6 in \cite{SangiorgiThesis}
but simplified to avoid any use of induction on type as appeared there.
\begin{lemma}
\label{lemma:beta-eq}
If $\cenv \types \app{\abst{\vx:\typet}{\pp}}{\valw}$ then
$\cenv \sts \app{\abst{\vx:\typet}{\pp}}{\valw} \parctxteq \pp [ \valw / \vx ]$.
\end{lemma}
In the following we will make use of a ``bisimulation up to'' argument \cite{SM92:ProblemWeak}.
\begin{definition}
A type-indexed relation $\rel$ is reduction-closed up to $({=},{\parctxteq})$
whenever $\cenv \sts \pp \rel \pq$ and $\pp \redn \pp'$ implies there exists some $\pq'$ such
that $\pq \Redn \pq'$ and $\cenv \sts \pp' \rel \parctxteq \pq'$.  
\end{definition}

\begin{lemma}
\label{lemma:upto}
For any type-indexed relation $\rel$ which is symmetric, reduction-closed up to
 $({=},{\parctxteq})$, $\Ppar$-contextual and barb-preserving, ${\rel} \subseteq {\parctxteq}$.
\end{lemma}

\begin{definition}
We say that $\vx$ is (un)guarded in $\pp$ whenever:
\begin{enumerate}
\item if $\vx \not\in \pp$ then $\vx$ is (un)guarded in $\pp$,
\item if $\vx \not\in \valw$ then $\vx$ is unguarded in $\app{\vx}{\valw}$,
\item if $\valv \neq \vx$ then $\vx$ is guarded in $\app{\valv}{\valw}$,
\item $\vx$ is guarded in $\inp{\valv}{\vy:\typet}{\pp}$, $\out{\valv}{\valw}{\pp}$,
and $\ifexp{\valv}{\valw}{\pp}{\pq}$, and 
\item if $\vx$ is (un)guarded in $\pp$ and $\pq$ then $\vx$ is (un)guarded in $\newnt{\ca}{\typet}{\pp}$,
$\pp \Ppar \pq$ and $\repl{\pp}$. \boxHere
\end{enumerate}
\end{definition}

\begin{lemma}
\label{lemma:parametricity}
  For any $\cenv, \vy: \tabs{\typet} \types \pr$ with $\vy$ guarded in $\pr$ 
and for any $\cenv \types \valv : \tabs{\typet}$ and   $\cenv \types \valw : \tabs{\typet}$,
if $\pr[\valv /\vy] \redn \pr'$ then 
$\pr' = \pr''[\valv/\vy]$ for some $\pr''$ and moreover,
$\pr[\valw/\vy] \redn \pr''[\valw/\vy]$.  
\end{lemma}
\proof%\begin{pf}
 We first observe that as $\cenv \types \valv : \tabs{\typet}$ it must be the case that 
$\valv$ is an abstraction and not a channel name. From this it is routine to 
check that the required property holds for the reduction axioms. Furthermore, if $\vy$ is
guarded in $\eval [ \pp ]$ then $\vy$ is guarded in $\pp$ and so 
the required property is preserved by reduction in evaluation contexts.
\qed%\end{pf}

\begin{lemma}
\label{lemma:guard-splitting}
  For any $\pp$ and $\vx$ we can find $\pq$ and $\vy$ such that $\vx$ is guarded in $\pq$,
$\vy$ is unguarded in $\pq$ and $\pp = \pq [\vx/\vy]$. 
\end{lemma}
\proof%\begin{pf}
 A routine induction on $\pp$.
\qed%\end{pf}

\begin{sloppy}
\begin{lemma}[Unguarded Substitutivity]
\label{lemma:substitutivity-unguarded}
  If $\cenv, \vx :\typet \sts \pp \parctxteq \pq$ and $\cenv, \vy : \tabs{\typet} \types \pr$
 and $\vy$ is unguarded in $\pr$ then $\cenv \sts \pr [\abst{\vx:\typet}{\pp} /\vy ] \parctxteq 
 \pr [\abst{\vx:\typet}{\pq} /\vy]$.
\end{lemma}
\end{sloppy}
\proof%\begin{pf}
  We proceed by induction on the structure of $\pr$.
If $\vy \not\in \pr$ then the result is immediate.
If $\pr$ is not of the form $\app{\valv}{\valw}$, the result follows easily by induction by making use of 
Lemma~\ref{lemma:cong}. Otherwise,  since $\vy$ is unguarded in $\pr$ we must have that
$\pr$ is of the form $\app{\vy}{\valw}$
with $\vy \not\in \valw$. Hence:
\[
\begin{array}{rcll}
\cenv \sts \pr [\abst{\vx:\typet}{\pp} /\vy] &=&
 \app{\abst{\vx:\typet}{\pp}}{\valw} & \mbox{(as $\pr = \app{\vy}{\valw}$ and $\vy \not\in \valw$)} \\
 & \parctxteq & \pp [\valw/\vx] & \mbox{(by Lemma~\ref{lemma:beta-eq})} \\
 & \parctxteq & \pq [\valw/\vx] & \mbox{(by hypothesis)} \\
 & \parctxteq &  \app{\abst{\vx:\typet}{\pq}}{\valw} & \mbox{(by Lemma~\ref{lemma:beta-eq})} \\
 & = & \pr [\abst{\vx:\typet}{\pp} /\vy] & \mbox{(as $\pr = \app{\vy}{\valw}$ and $\vy \not\in \valw$)}. 
\end{array}
\]
as required.
\qed%\end{pf}

\begin{sloppy}
\begin{lemma}[Guarded Substitutivity]
\label{lemma:substitutivity-guarded}
  If $\cenv, \vx :\typet \sts \pp \parctxteq \pq$ and $\cenv, \vy : \tabs{\typet} \types \pr$
 and $\vy$ is guarded in $\pr$ then $\cenv \sts \pr [\abst{\vx:\typet}{\pp} /\vy ] \parctxteq 
 \pr [\abst{\vx:\typet}{\pq} /\vy]$.
\end{lemma}
\end{sloppy}
\proof%\begin{pf}
  Let $\rel$ be defined as 
\[ 
\cenv \sts \pr' [\abst{\vx:\typet}{\pp} /\vy ] \rel
 \pr' [\abst{\vx:\typet}{\pq} /\vy] \mbox{ whenever } \cenv, \vy:\tabs{\typet} \types \pr'
 \mbox{ and } \vy \mbox{ is guarded in } \pr'
\]
We show that $\rel$ is symmetric, reduction-closed up to $({=},{\parctxteq})$, $\Ppar$-contextual,
and barb-preserving and so the result follows by Lemma~\ref{lemma:upto}.
Symmetry, $\Ppar$-contextuality, and barb-preservation are direct. 
For reduction-closure up to $({=},{\parctxteq})$ we suppose:
\[
  \pr' [\abst{\vx:\typet}{\pp} /\vy] \redn \pr''
\]
By Lemma~\ref{lemma:parametricity} we have that
$\pr'' = \pr'''[\abst{\vx:\typet}{\pp} /\vy]$ and moreover:
\[
  \pr' [\abst{\vx:\typet}{\pq} /\vy] \redn \pr'''[\abst{\vx:\typet}{\pq} /\vy]
\]
We use Lemma~\ref{lemma:guard-splitting} to find a $\pr''''$ and $\vz$ such that
$\vy$ is guarded in $\pr''''$, $\vz$ is unguarded in $\pr''''$ and $\pr'''=\pr''''[\vz/\vy]$.
Hence:
\[\begin{array}{rcll}
  \pr'' & = & \pr'''[\abst{\vx:\typet}{\pp} /\vy] & \mbox{(from above)} \\
        & = & \pr''''[\abst{\vx:\typet}{\pp} /\vy,\abst{\vx:\typet}{\pp} /\vz] & \mbox{(from above)} \\
        & \rel & \pr''''[\abst{\vx:\typet}{\pq} /\vy,\abst{\vx:\typet}{\pp} /\vz] & \mbox{(from definition of $\rel$ and $\vy$ guarded in $\pr''''[\abst{\vx:\typet}{\pp} /\vz]$)} \\
        & \parctxteq & \pr''''[\abst{\vx:\typet}{\pq} /\vy,\abst{\vx:\typet}{\pq} /\vz] & \mbox{(from Lemma~\ref{lemma:substitutivity-unguarded} and $\vz$ unguarded in $\pr''''[\abst{\vx:\typet}{\pq} /\vy]$)} \\
        & = & \pr'''[\abst{\vx:\typet}{\pq} /\vy] & \mbox{(from above)}
\end{array}\]
as required.
\qed%\end{pf}

\begin{corollary}
\label{corollary:substitutivity}
 If $\cenv, \vx :\typet \sts \pp \parctxteq \pq$ and $\cenv, \vy : \tabs{\typet} \types \pr$
 then $\cenv \sts \pr [\abst{\vx:\typet}{\pp} /\vy ] \parctxteq 
 \pr [\abst{\vx:\typet}{\pq} /\vy]$.
\end{corollary}
\proof%\begin{pf}
  Follows from Lemmas~\ref{lemma:guard-splitting},~\ref{lemma:substitutivity-unguarded} 
and~\ref{lemma:substitutivity-guarded}.
\qed%\end{pf}

\begin{corollary} \label{corollary:cong-hard} \mbox{}
  \begin{enumerate}
  \item  If $\cenv, \vx:\typet \sts \pp \parctxteq \pq$ and $\cenv \types \valv : \tabs{\typet}$
then $\cenv \sts \app{\valv}{\abst{\vx:\typet}{\pp}} \parctxteq \app{\valv}{\abst{\vx:\typet}{\pq}}$.
  \item  If $\cenv, \vx:\typet \sts \pp \parctxteq \pq$, $\cenv \types \ca : \tchan{\tabs{\typet}}$
and $\cenv \types \pr$
then $\cenv \sts \out{\ca}{\abst{\vx:\typet}{\pp}}{\pr} \parctxteq \out{\ca}{\abst{\vx:\typet}{\pq}}{\pr}$.
  \end{enumerate}
\end{corollary}
\proof%\begin{pf}
  Follows from Corollary~\ref{corollary:substitutivity}.
\qed%\end{pf}

\begin{pf}[of Lemma~\ref{lemma:context-lemma}]
 The `only if' direction is immediate. For the converse it is sufficient to
 show that $\parctxteq$ is preserved by each process operator of \hopi~as demonstrated
by Lemma~\ref{lemma:cong} and Corollary~\ref{corollary:cong-hard}.
\end{pf}

\section{Merge is a partial function}
\label{app:mg-ok}

\begin{pf}[of Proposition~\ref{prop:mg-ok}]
We consider the rewriting relation $\rgrewrite$ which we will define
as the one-step rewriting used to define the merge operation:
  \begin{axioms}
     \conf  & \rgrewrite & \checkmark \quad &  
    \mbox{if } \conf \mbox{ doesn't contain } \res{\tnk}{\valv} \mbox{
      for any } \tnk, \valv \\
%\mbox{ implies } \tnk \not\in \kenv 
    \newnt{\vec\ca}{\vec\typet}{\res{\tnk}{\valv} \Ppar
        \conf} & \rgrewrite &                           
                          \newnt{\vec\ca}{\vec\typet}{\conf[\valv / \call{\tnk}]} & 
    \mbox{if } \call{\tnk} \not\in \valv \\
  \end{axioms}
It is easy to see that $\rgrewrite$ is a terminating rewriting
relation.  Moreover, the rewriting will terminate with a $\checkmark$  from 
$\conf$ (so that $\mg{\conf}$ is defined)
exactly when $\rg{\conf}$ is acyclic.
To see this we consider the effect of $\rgrewrite$ on reference
graphs: for 
$$
  \res{\tnk}{\valv} \Ppar
        \conf   \qquad \rgrewrite \qquad \conf[\valv / \call{\tnk}]
$$
the reference graph of $\res{\tnk}{\valv} \Ppar \conf$ has the node
$\tnk$ removed and any edges such that
$$
  \tnl' \mapsto \tnk \mapsto \tnl 
$$
for $\tnl', \tnl \neq \tnk$, 
are replaced with an edge
$$
 \tnl' \mapsto \tnl
$$
all other edges involving $\tnk$ are removed.
So if node $\tnk$ is involved in a cycle before rewriting occurs,
that is 
$$
\tnl \trans{\mapsto} \tnk \trans{\mapsto} \tnl
$$ 
for some $\tnl$, then either it is a
\emph{tight loop}, that is $\tnl = \tnk$ and $\tnk \mapsto \tnk$, or
$\tnl \neq \tnk$ and the cycle still
exist after rewriting as $\tnl \trans{\mapsto} \tnl$. 
The side-condition on the rewrite rule forbids
tight loops hence we see that $\rgrewrite$ preserves cyclicity. That
is: 
\begin{quote}
if $\conf \rgrewrite \conf'$ then $\rg{\conf}$ is acyclic if and only
if $\rg{\conf'}$ is acyclic.
\end{quote}
Now, suppose that $\mg{\conf}$ is defined. We know that there exists a
finite sequence
$$
\conf \rgrewrite \conf_1 \rgrewrite \cdots \rgrewrite \conf_n \rgrewrite \checkmark
$$
with $\mg{\conf} = \conf_n$. 
We know that $\rg{\conf_n}$ is acyclic as it contains no edges. Thus,
$\rg{\conf}$ is acyclic also.
Conversely, suppose that $\rg{\conf}$ is acyclic. Then as $\rgrewrite$
is terminating there must be a finite sequence 
$$
\conf \rgrewrite \conf_1 \rgrewrite \cdots \rgrewrite \conf_n
$$
such that $\conf_n$ cannot be rewritten. 
There are two possibilities for this: either $\rg{\conf_n}$ contains a
tight loop, or $\conf_n$ is $\checkmark$. We see that $\rg{\conf}$ is
acyclic, so $\conf_n$ is acyclic too and therefore cannot contain a
tight loop. Thus $\conf_n$ is $\checkmark$ and $\mg{\conf}$ is defined.

To show that $\mg{\cdot}$ is a well-defined partial function it
suffices to show that it is strongly confluent for acyclic terms. 
Note that if $\newn{\ca:\typet}{\conf} \rgrewrite \conf'$ then either
$\conf'$ is $\checkmark$ or $\conf' \equiv \newn{\ca:\typet}{\conf''}$ such
that $\conf \rgrewrite \conf''$. So without loss of generality 
suppose that 
$$\conf \rgrewrite \conf_1 \qquad  \mbox{and}  \qquad \conf \rgrewrite \conf_2$$
for 
$$\conf \equiv \conf_1' \Ppar \res{\tnk_1}{\valv_1} \qquad \mbox{and}
\qquad \conf \equiv \conf_2' \Ppar \res{\tnk_2}{\valv_2}$$ 
so that 
$$
\conf_1 \equiv \conf_1'[\valv_1/\call{\tnk_1}]
\qquad
\mbox{and}
\qquad
\conf_2 \equiv \conf_2'[\valv_2/\call{\tnk_2}].
$$
So either, $\tnk_1 = \tnk_2$ in which case $\conf_1 \equiv \conf_2$ or
$\tnk_1 \neq \tnl_2$ and 
$$
\conf_1' \equiv \conf_3' \Ppar \res{\tnk_2}{\valv_2}
\qquad 
\mbox{and}
\qquad
\conf_2' \equiv \conf_3' \Ppar \res{\tnk_1}{\valv_1}
$$
We notice that 
$$
\begin{array}{llcl}
& \conf_1 & \equiv & \conf_1'[\valv_1/\call{\tnk_1}] \\
&         & \equiv & (\conf_3' \Ppar \res{\tnk_2}{\valv_2})
[\valv_1/\call{\tnk_1}] \\
&         & \equiv & \conf_3'[\valv_1/\call{\tnk_1}] \Ppar
\res{\tnk_2}{\valv_2[\valv_1/\call{\tnk_1}]} \\
(\mbox{acyclicity implies } \call{\tnk_2}
\not\in\valv_2[\valv_1/\call{\tnk_1}]) &         & \rgrewrite &
\conf_3'[\valv_1/\call{\tnk_1}][\valv_2[\valv_1/\call{\tnk_1}] /
\call{\tnk_2}] \\
 &        & \equiv & \conf_3'[\valv_1[\valv_2[\valv_1/\call{\tnk_1}] /
\call{\tnk_2}] / \call{\tnk_1} , \valv_2[\valv_1/\call{\tnk_1}] /
\call{\tnk_2} ] \\
(\mbox{acyclicity})  &        & \equiv &
 \conf_3'[\valv_1[\valv_2/\call{\tnk_2}]/\call{\tnk_1}, \valv_2[\valv_1/\call{\tnk_1}] /
\call{\tnk_2}] \\
 (\mbox{def}) &  & \equiv & \conf_3
\end{array}
$$
By a symmetric argument we see that 
$\conf_2 \rgrewrite \conf_3'[\valv_2[\valv_1/\call{\tnk_1}]/\call{\tnk_2}, \valv_1[\valv_2/\call{\tnk_2}] /
\call{\tnk_1}]$ and, by definition, this is just $\conf_3$ so we have 
$\conf_2 \rgrewrite \conf_3$. 
Thus $\rgrewrite$ is strongly confluent for acyclic terms
and hence $\mg{\cdot}$ is well-defined.
\end{pf}

\bibliographystyle{plain}
\bibliography{hopi}

\end{document}

%%% Local Variables: 
%%% mode: latex
%%% TeX-master: t
%%% End: 